# In-situ real-time evolution of intrinsic stresses and microstructure during growth of cathodic arc deposited (Al,Ti)N coatings

Sanjay Nayak[1], Tun-Wei Hsu[1], Lina Rogström[1], Maiara Moreno[1], Jon M. Andersson[2], Mats P. Johansson-Jöesaar[1,2], Robert Boyd[1], Norbert Schell[3], Jens Gibmeier[4], Jens Birch[1], and Magnus Odén[1]

[1]Department of Physics, Chemistry and Biology (IFM), Linköping University, SE-581 83 Linköping, Sweden
[2] Seco Tools AB, 737 82, Fagersta, Sweden
[3]Helmholtz-Zentrum Hereon, Institute of Materials Physics, Max-Planck-Str. 1, Geesthacht 21502, Germany
[4]Institute for Applied Materials, Materials Science and Engineering (IAM-WK), Karlsruhe Institute of Technology, 76131, Karlsruhe, Germany

**Abstract**

The residual stress plays a vital role in determination of the device performance that uses thin films coating and thus the accurate determination of stress and its optimization with process parameters is an ongoing research work for many decades. In line with this, the microscopic origin of the stress at the atomic scale and its development during the thin film deposition is a matter of major scientific interests. The development of stress is a complex phenomenon and has a complex dependence to process parameters, film microstructure and its morphology. In this work, by utilizing a custom-designed cathodic arc deposition system and synchrotron radiation based 2D x-ray diffraction (XRD) technique, we determine the real-time evolution of stress, crystallite sizes and their preferential orientations of Aluminum-Titanium-Nitride ($Al_xTi_{1-x}N$) films with varied Al-content (x=0.0, 0.25, 0.50, and 0.67) on Si-100 substrate. The energies of incoming ions and hence stress in the films is tuned by applying different direct current substrate bias ($V_s$ = floating potential, -20, -40, -60, -80, and -100 V). The instantaneous stress is evaluated by the well-known d vs. $\sin^2\psi$ technique, while crystallite sizes are determined by analyzing line profiles of x-ray diffractograms. The evolution of stress and crystallite sizes are modelled with multiple numerical models from which kinetic parameters associated with the thin film depositions are extracted. The ex-situ microstructure characterizations of $Al_xTi_{1-x}N$ coatings are carried out by scanning electron microscopy (SEM) and transmission electron microscopy (TEM). The formation of ex-situ microstructure of the films is discussed considering the results obtained from in-situ XRD data. Finally, we demonstrate that the method utilized here is a powerful approach towards estimation of the fracture toughness of thin film coatings.

## I. INTRODUCTION:

Thin film coatings are the building blocks of today's technology. Their applications are widespread, such as in, semiconductor devices [1,2], optoelectronics [3], micro and nano electromechanical systems [4], detectors [5], energy storage [6], optical mirrors [7], and wear and corrosion-resistant coating [8], etc. Since thin film are deposited over a lattice-mismatched surface of a single crystalline substrate material, they are often strained. In addition, the deposition process is a non-equilibrium phenomenon, leading to the generation of residual stresses in the coatings [9]. This residual stress plays a vital role in the performance during their application. While a high value of tensile stress leads to cracking [10], compressive stress can cause buckling, blistering, and delamination of the coatings [11–13]. Thus, accurate determination and optimization of stress in the thin film is of great technological importance, and hence over the last 100 years it has been pursued [14,15]. The strain/stress is also an important external parameters that can enabled many material properties, such as electrical and thermal conductivity [16], ferroelectricity [17] , piezoelectricity [18], optical properties [19], mechanical hardness [20,21], etc. Besides their post-deposition stress determination, understanding the evolution of stress during their deposition is of scientific interest.

There are various methods for the determination of residual stresses in coatings, such as wafer curvature measured by multibeam optical stress sensors (MOSS) [9], synchrotron x-ray diffraction [22], holographic transmission electron microscopy (TEM) [23], micro-Raman spectroscopy [24], and electron diffraction [25]. Amongst all, largely MOSS has been successfully used in the in-situ evaluation of stress during thin film depositions. While MOSS is very powerful in determining the macroscopic stress of the entire films for both crystalline as well as amorphous samples, it will not reveal any information about microstructure evolution. In contrast, x-ray diffraction can provide both in-situ strain and information on microstructure simultaneously, but it is limited to only crystalline materials. Using MOSS, the study of stress evolution during the growth of polycrystalline thin films is extensively investigated, but most of them are on pure metal films on insulator substrates only [26–29]. In addition, a few reports are available where MOSS was used to study the stress evolution in a complex compound system, such as transition metal nitrides [30–33]. However, all these nitrides thin films were deposited with a relatively lower deposition rate technique, such as magnetron sputtering.

Polycrystalline thin film of multi-component alloys of transition metals nitrides (TMNs), particularly (Al,Ti)N-based coatings are the current workhorse of the cutting tool industry. The success of the (Al,Ti)N materials system compared to traditional TiN hard coatings has been attributed to 1) increased oxidation resistance with increasing aluminium content [34]; and 2) the associated age-hardening effect with elevating working temperature, due to the formation of coherency strains between c-TiN and metastable c-AlN. domains by spinodal decomposition [35]. For superior oxidation resistant and enhanced thermomechanical properties, a higher concentration of Al is preferable, but the aluminium content in cubic-(Al,Ti)N is limited to ≈ 65% due to thermodynamic limitations [36,37]. Above 65% of Al content, the formation of additional hexagonal (Al,Ti)N crystallizes, which has inferior mechanical properties compared to the cubic phase [38,39]. The mechanical properties and consequently the tool's performance is also influenced by the microstructure of the coating, i.e., the grain size, density, grain orientation, dislocation density, etc. Suggested by the Hall-Petch relation[40], that above a critical grain size the hardness of the overall coating material increases with decreasing grain size. A common approach to modify the grain size can be accomplished by energetic ion bombardments, which are controlled by applying a negative electrical bias to the substrate surface [39,41]. The physical vapour depositions process especially in cathodic arc evaporation comprises a high density of metal ions and multiple charge states, thus their kinetic energy can be increased substantially by applying a negative DC bias to the substrate [42]. Upon bombardment of these high-speed ions on top of already growing surfaces of the films can modify the stress level as well as the orientation of the grains in the polycrystalline thin films[43,44]. Such a change in the preferred orientation strongly correlates to its mechanical properties, [45–49] hence also its performance during the actual service conditions [50,51].

Although cathodic arc deposition is the industrial preferred technique for (Al,Ti)N-based coatings, the evolution of microstructure and strain/stress during deposition are still not well understood and scarcely studied. Here we present the in-situ real-time evolution of coating stress and microstructure during growth of ultra-high vacuum (UHV) cathodic arc deposited (Al,Ti)N alloy coatings as a function of substrate bias ($V_s$). Both the stress state and microstructure of the coatings were evaluated and analysed by synchrotron x-ray diffractograms recorded during film growth. All depositions were carried out in a specially designed UHV deposition system, adapted for synchrotron radiation studies [52].

## II. EXPERIMENTAL DETAILS:

Alloyed (Al,Ti)N coatings with varied Al-content ($Al_xTi_{1-x}N$) on Si (100) substrate are deposited by reactive cathodic arc deposition technique, where the deposition system is schematically shown in Figure 1. 63 mm diameter size metallic Ti (Grade-2), and alloys of $Al_{0.25}Ti_{0.75}$, $Al_{0.50}Ti_{0.50}$, and $Al_{0.67}Ti_{0.33}$ (FK-Grade) are used as cathodes. The cathode current is fixed at 75 A for all depositions.

The cathodes are mounted on home-designed holders, which are further mounted on a specially designed chamber lid [52]. The angle between the surface normal of the cathodes to the substrate normal is 35° (see Figure 1). The distance between the substrate and the cathode is ≈ 12 cm. Pure (99.9995%) nitrogen ($N_2$) flow is introduced into an ultra-high vacuum system with a base pressure of $10^{-8}$ Torr. During the deposition of all $Al_xTi_{1-x}N$, the growth pressure is fixed at 40 mTorr. Prior to deposition the Si (100) substrate is cleaned by ultrasonic bath with isopropyl alcohol and dried by blowing dry air. For all depositions, the substrate is heated up to 450°C by an 808 nm infrared laser diode (Dilas Diode Laser, Inc.) powered by a 300 W power supply (Lumina Power, Inc.). The laser beam is focused on a susceptor material, here silicon carbide (SiC), which then heats the substrate via thermal conduction and radiation. The substrate is rotated at a speed of 5 rotations per minute (RPM). The applied volage in electrical bias to substrate ($V_s$) is applied by using a DC power supply, and is varied between the depositions from -20 to -100 V in steps of -20 V. In addition, depositions at floating potential (FP) are also performed. The duration of the deposition of all coatings is fixed at 10 minutes. During the experiments, the deposition chamber was fixed to a high-resolution hexapod with a resolution in x/y/z movement of ±1 μm.

The film microstructure and final thickness is determined from a fractured cross-sectional view using a FESEM (Leo 1550 Gemini) operated at 5 kV. Microstructural characterization of the coatings was made by TEM and STEM using an FEI Tecnai G220 UT microscope operated at 200 kV. In-situ x-ray diffractograms of $Al_xTi_{1-x}N$ coatings are recorded in transmission mode during the deposition of the coatings by utilizing synchrotron radiation at the P07 High Energy Materials Science (HEMS) beamline at PETRA III (Deutsches Elektronen-Synchrotron (DESY); Hamburg, Germany). The energy of the x-ray photons is selected to 73.79 keV. The beam slit widths are fixed at 100×600 μm². The diffractograms are recorded with a 2D flat panel detector with 2048x2048 pixles (PerkinElmer XRD 1622). The exposure time for recording one diffractogram is 0.2 s, and ten such patterns are superposed with each other to generate a final diffractogram. The distance between substrate and detector is evaluated by recording the diffractogram of NIST standard $LaB_6$ powder. In this set of experiments, the distance is determined to be 3.373 m. The detector was positioned to collect one quadrant of the diffraction pattern for optimizing the resolution.

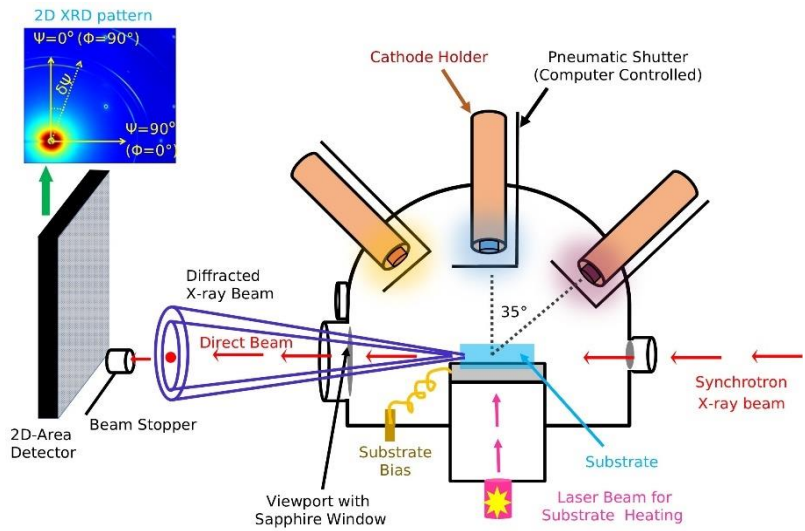

*Figure 1. The sketch of the experimental setup of our experiments. The angle between substrate's surface normal and the normal to cathode targets surface is 35°. The relationship between Φ and ψ is also shown in the 2D- XRD pattern.*

The 2-dimensional XRD patterns are transformed into one-dimensional lineouts by integration in ≈ 5°-8° wide azimuthal bins ($\psi$). The interplanar lattice spacing ($d_{hkl}$) is determined by fitting the one-dimensional lineout with a pseudo-Voight function. The details of the method used in estimation of in-plane strain and biaxial stress is given in section SI. I of supplementary information (SI). In determination of stress, high temperature (450 °C) elastic constants of $Al_xTi_{1-x}N$ are used (see Ref. [53] and section SI. I of SI for details).

The crystallite size is approximated to coherently diffracting domains ($\langle D_{hkl}\rangle$), and at a film thickness, $t_f$, ($\langle D_{hkl}(t_f)\rangle$) is estimated using the Debye-Scherrer formula [54].

$$\langle D_{hkl}(t_f = t)\rangle = \frac{0.9 \times \lambda}{\zeta(t_f = t)\cos\theta(t_f = t)} \quad (1)$$

Where $\lambda$ is the wavelength of the x-ray beam, $\zeta(t_f = t)$ and $\theta(t_f = t)$ are the the instrumental broadening corrected FWHM of the one-dimensional lineouts in radians and half of diffraction angle, $2\theta$, of $hkl$ crystal planes at $t_f = t$, respectively. The instrumental broadening was determined from the FWHM of the diffraction signal from the LaB$_6$ powder. The $\langle D_{hkl} \rangle$ of Al$_x$Ti$_{1-x}$N coatings along the GD ($D_{111}^{GD}$) as deposited with different V$_s$ are obtained by integrating the intensity of diffractograms from ψ = 0° to 10° of c-111 crystal planes. Evolutions of D$_{111}$ along the IP-direction ($D_{111}^{IP}$) are obtained by measuring line width profiles of the diffractograms formed close to the IP-direction of c-111 crystal planes. The respective values of range of ψ used in integration of diffraction intensities are mentioned in the later sections.

The evolution of average crystallite size, $\langle D_{hkl}(t_f) \rangle$ with the film thickness $t_f$, along the in-plane direction is modelled with a power law as follows.

$$\langle D_{hkl}(t_f) \rangle = k \times t_f^n \quad (2)$$

The evolution of film thickness, $t_f$, average $\sigma$ as a function of $t_f$ is modelled using a power law,

$$\langle \sigma(t_f) \rangle = \sigma_0 + c \times t_f^{-\gamma} \quad (3)$$

The $\sigma_0$ are the converged $\langle \sigma(t_f) \rangle$ at large $t_f$. The scaling parameter $c$ and $\varepsilon'$ are the coefficients related to the rate of grain boundary shrinkage with an exponent of $\gamma$ [55]. The $\sigma$ evolution of Al$_x$Ti$_{1-x}$N coatings is also modelled using a recently developed kinetic model proposed by Chason *et al.* [56] and is given by:

$$\langle \sigma(t_f=t) \rangle = \sigma_c + \left( \frac{\sigma_{T,0}(L_0)}{L(t_f=t)^{0.5}} - \sigma_c \right) \times \exp(-\beta D_{eff}/RL(t_f=t)) + A\left(\frac{1}{L(t_f=t)}\right) + \left(1 - \frac{1}{L(t_f=t)}\right) \times B \times \frac{1}{\left(1 + \frac{1}{R\tau_s}\right)} \quad (4)$$

where $\sigma_c$ is the compressive stress that is generated by the insertion of ions at the grain boundary, which is fixed at -0.02 GPa in our work [30]. $\sigma_{T,0}(L_0)$ is the tensile stress generated by the formation of a new segment of grain boundaries with a grain size of L$_0$. $L_{(t_f=t)}$ is the average grain size at $t_f = t$, approximated to in-plane $\langle D_{hkl}(t_f) \rangle$ in our fitting. $\beta$ is a parameter that depends upon the mechanical properties of the layer and concentration of mobile defects. $D_{eff}$ is the effective diffusivity of atoms from surface to grain boundary. R is the rate at which grain boundary height is changing, which is approximated by the film growth rate in our work. The third and fourth terms of Eqn. 4 represent stress due to energetic vapour fluxes. The third term is related to collision-induced densification of grain boundary, while the last term is due to the formation of defects in bulk with a depth of $l$ from the surface. Parameter $\tau_s$ is the characteristic time at which defects created in bulk migrate to the surface and annihilate. The parameter $\tau_s$ is further approximated to,

$$\tau_s = \frac{l}{R}\left[\alpha - 1 - \alpha\sqrt{1 - \frac{2}{\alpha}}\right]$$

where $\alpha = D_i/2Rl$. $D_i$ represent the diffusivity of the defect created in bulk. Parameters A and B are adjustable parameters. In this work, a least-squares method is used to fit Eqn.3 and 4 to the experimentally obtained data. While fitting of Eqn. 7 is named as Model 1, fitting of experimental $\sigma$ with Eqn. 8 will be named as Model 2 in the rest of this paper. In fitting of Model 2 to the experimental data, A and B are restricted to have a negative value. While $\sigma_c$ is fixed at -0.2 GPa for all compositions of Al$_x$Ti$_{1-x}$N, $\sigma_{T,0}(L_0)$ is tuned to get a better fitting, but kept fixed for one composition, i.e., 0.67 GPa, 0.85 GPa, 1.00 GPa, and 2.50 GPa for L$_0$ = 100 nm in TiN, Al$_{0.25}$Ti$_{0.75}$N, Al$_{0.50}$Ti$_{0.50}$N, Al$_{0.67}$Ti$_{0.33}$N, respectively. Quality of fitting of both Model 1 and Model 2 are assessed by a R-squared ($R^2$) analysis.

### III. FORMATION AND EVOLUTION OF MICROSTRUCTURE:

#### A. Formation of crystal phases, and crystallographic texture

Recorded diffractograms of all samples at the end of deposition are displayed in Figure 2. The corresponding Miller indices of different diffraction spots/rings are identified by measuring interplanar distances (or corresponding 2θ values) and are marked accordingly. $Al_xTi_{1-x}N$: $x \leq 0.5$ coatings show the formation of only cubic crystal phases (*c*-) (see Figure 2 (a-o)), while $Al_{0.67}Ti_{0.33}N$ coatings grown with $|V_s| \leq 60$ V show mixed cubic and hexagonal phases (*h*-) (see Figure 2 (p-r)). At a higher $V_s$ (i.e., -80 and -100 V) no detectable hexagonal phases of $Al_xTi_{1-x}N$ are seen (see Figure 2 (s-u) and Figure S-6 (a) of SI).

TiN coating deposited at $V_s$ = FP shows a preferred orientation of c-111 planes along the growth direction (GD), along with a minority component at an angle of ψ ≈ 74° to the growth direction (see Figure 2 (a)). The angle between the (111) and (200) directions in a cubic crystal is ~55°, corresponding well to the preferred orientation seen for the c-200 rings (see Figure 2 (a)). TiN coatings deposited with $V_s$ = -20V and -40 V, the crystallites in the films orient more randomly, but higher intensity diffraction spots are present in c-200 ring at an angle of ψ ≈ 22°/20° and ψ ≈70°/77° (see Figure 2 (b-c)). TiN samples deposited at $V_s$ = -60, -80, and -100 V show the formation of the preferred orientation of crystallites with c-111 along the GD and intense diffraction from c-200 appears close to a 18˚~20° inclination angle to the GD (see Figure 2 (d-f)). The texture of c-200 at close to ψ ≈70° is consequent of the 90° rotational symmetry of the c-200 crystal planes in the FCC lattice.

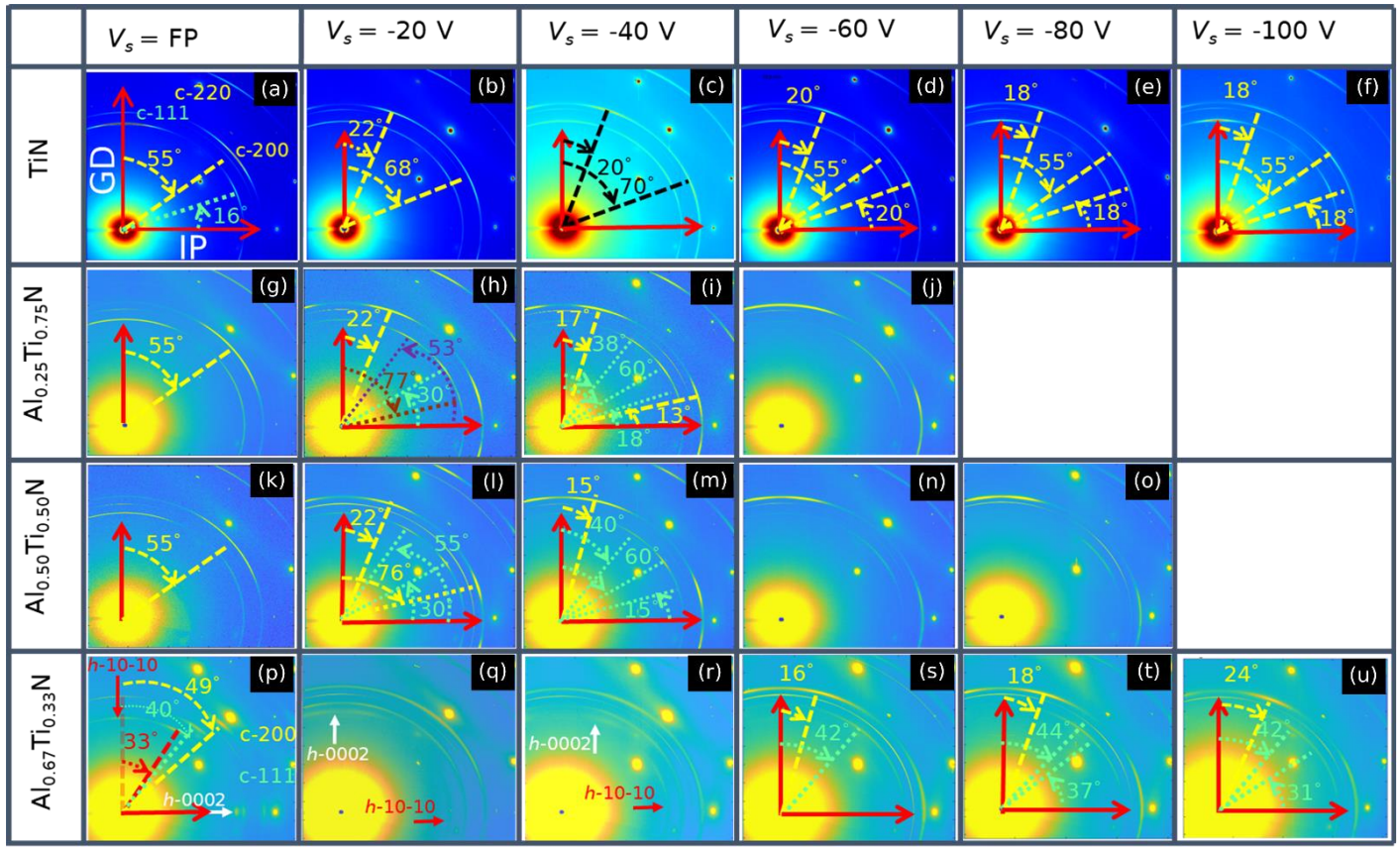

*Figure 2. 2D x-ray diffractograms of $Al_xTi_{1-x}N$ coatings deposited on Si (100) substrate at different substrate bias voltages ($V_s$). Diffraction from hkl planes of cubic phases is denoted as c-hkl, while for hexagonal phases they are denoted as h-hkil where i=-(h+k). The unmarked high intensities spots are from Si substrates.*

The $Al_{0.25}Ti_{0.75}N$ coating deposited at $V_s$ = FP shows a preferred orientation of c-111 along the GD and c-200 at an angle ψ ≈55 (see Figure 2 (g)). With $V_s$ = -20 V, the orientation of c-200 planes splits along two distinct ψ orientations: (i) one at ψ ≈22° and (ii) at ψ ≈77° (see Figure 2 (h-i)). There are three distinct intense regions in the c-111 Debye ring: (i) at ψ ≈ 37°, (ii) at ψ ≈ 60°, and (iii) at ψ ≈77°. The application of $V_s$ also influenced the texture of c-111 crystallographic planes. The diffractograms of $Al_{0.25}Ti_{0.75}N$ coatings for both $V_s$ = -20 and -40 V are similar, except in the latter case the c-200 crystal planes are tilted a bit closer towards the GD with a tilt angle of 17°. The intense spots of c-111 at an angle ψ ∈ [37°, 40°] are from the crystallographic c-111 planes of the grains which c-200 planes are inclined at angle of ψ ∈≈ [-15°, -20°]. The intense spots of c-111 at an angle ≈ 60° are from the same grains for which c-200 planes are at an inclined angle close to ψ ∈ [+15°, +20°]. The intense c-111 close ψ ≈ 75° is from different grains altogether, whose c-200 diffraction spots should be close to ψ ≈45°, i.e., tilted away from the GD. A faint c-200 in these regions indicate the grain volume of these grains are smaller. With further increase in $V_s$ to -60 V the c-200 crystal planes are found to be along the GD with a broader azimuthal distribution (see Figure 2 (j)). Note that the $Al_{0.25}Ti_{0.75}N$ coating deposited with $V_s$ = -60 V shows a cohesive failure to the substrate during its film deposition. Similar behaviour in the diffractogram of $Al_{0.50}Ti_{0.50}N$ is also seen. The coatings deposited at $V_s$ = -60 V and -80 V both show a predominant texture of c-200 along the GD and cohesive failure of the coatings (discussed later) (see Figure 2 (n-o)).

Unlike the above-discussed compositions of $Al_xTi_{1-x}N$: x ≤0.5, the diffractograms from $Al_{0.67}Ti_{0.33}N$ coatings show a different crystallographic texture formation. Formation of both hexagonal and cubic phases crystals are noticed (see Figure 2(p-u) and Figure S-6 of SI). X-ray diffractograms recorded from $Al_{0.67}Ti_{0.33}N$ coating deposited with $V_s$ = FP show h-10$\bar{1}$0 crystal planes are aligned along the GD and at an inclined angle of 33° to the GD (see Figure 2 (p)). A clear and sharp diffraction spot corresponding to h-0002 planes is in parallel to the IP direction (see Figure 2 (p)). The c-111 and c-200 crystal planes of $Al_{0.67}Ti_{0.33}N$ are inclined at an angle 40° and 49° to the GD, respectively. Weakly textured c-111 and c-200 crystal planes of c-AlTiN phases are also seen along the IP (see Figure 2 (p)). Since the angle between these two c-111 and c-200 is not in agreement with the crystallographic angle (≈55°) between c-111 and c-200 in FCC lattice, we suggest both are from two distinct grains with different orientations. Based on analysis of symmetry of the crystal structure, we find that the c-200 crystal planes of the grains whose c-111 planes are at an angle of 40° to the GD should be at around 95° from the GD [hence 5° from the IP direction]. And the c-111 planes of grains whose c-200 are at inclined angle of 49° to the GD should be at around 84° from the GD. Presence of diffraction spots of c-111 and c-200 along and close to the IP direction agrees well with this prediction. With $V_s$= -20 and -40 V, the crystal planes' orientation changes dramatically. The crystallographic orientation of c-111 and c-200 planes are random, along with h-0002 and h-10$\bar{1}$0 planes oriented along the GD and towards the IP direction, respectively (see Figure 2 (q-r)). Surprisingly, for $Al_{0.67}Ti_{0.33}N$ coatings deposited with $|V_s| \geq 60$ V, no sign of diffractograms corresponding to hexagonal phase are observed (see Figure 2 (s-u)). The intense spots of the Debye ring corresponding to c-200 crystal planes of $Al_{0.67}Ti_{0.33}N$ are at an angle 16°, 18°, and 24° to the GD for coatings deposited at $V_s$= -60, 80, and 100 V, respectively. The c-111 planes are orientated at an angle of 42°/44° to the GD for all films (see Figure 2 (s-u)). A fixed tilt angle of intense of c-111 and varied c-200 clearly establish that both intense diffraction patterns are originated from different grains with different orientations.

### B. Ex-situ surface morphology, film thickness, and growth rate

Cross-section SEM micrographs revealed that all films have a columnar structure (see Figure 3). The total film thickness $t_f$ of $Al_xTi_{1-x}N$ coating, determined from the cross-sections viewed in the SEM, are tabulated in Table 1. The TiN coating grown at $V_s$ = FP shows the formation of relatively smaller domain size, and loosely packed nanocolumns with gaps (voids) between neighbour nanocolumns at the bottom part of the coating extending to the top. With an increase in $|V_s|$ (≥ 20 V) during the thin film deposition, the TiN coatings densify with a larger domain size of nanocolumns. A closer look into the cross-sectional SEM images of TiN coatings deposited with a finite $V_s$ reveals the formation of individual thin NCs until $t_f$ of ≈ 200-300 nm beyond which they coalescence to form broad nanocolumns (NCs).

Similar morphology evolution under different $V_s$ is observed for $Al_{0.25}Ti_{0.75}N$ and $Al_{0.50}Ti_{0.50}N$. A major difference in the cross-sectional morphologies of $Al_{0.25}Ti_{0.75}N$ and $Al_{0.50}Ti_{0.50}N$ coatings deposited at $V_s$ =FP is noticed where in the latter case the nanocolumns are more isolated than the former one. $Al_{0.25}Ti_{0.75}N$ and $Al_{0.50}Ti_{0.50}N$ coatings deposited above $|V_s| \geq 60$ V show characteristic features of a coating with cohesive failure that occurred during the film deposition. In these films two distinct $t_f$ can be seen (see Figure 3 (j, n, and o)). The electron micrographs obtained from $Al_{0.67}Ti_{0.33}N$ coating differ from those with lower Al content. While $Al_{0.67}Ti_{0.33}N$ films grown at $V_s$= FP a nanocolumnar structure is seen, with an applied $V_s$ the cross-sectional SEM image shows a densely packed nanocrystalline like morphology.

*Table 1 Total film thickness ($t_f$) of $Al_xTi_{1-x}N$ coatings deposited with different substrate bias voltage ($V_s$). The deposition time is fixed for 10 minutes.*

| $V_s$ (V) | $t_f$ (nm) | | | |
|---|---|---|---|---|
| | TiN | $Al_{0.25}Ti_{0.75}N$ | $Al_{0.50}Ti_{0.50}N$ | $Al_{0.67}Ti_{0.33}N$ |
| FP | 1007(±12) | 1631(±12) | 1640(±20) | 1867(±12) |
| -20 V | 917(±8) | 1494(±10) | 1545(±15) | 1573(±15) |
| -40 V | 920(±8) | 1508(±10) | 1508(±12) | 1677(±11) |
| -60 V | 912(±8) | 1471(±10) | 1545(±10) | 1615(±11) |
| -80 V | 930(±10) | - | 1520(±10) | 1508(±10) |
| -100 V | 895(±10) | - | - | 1502(±15) |

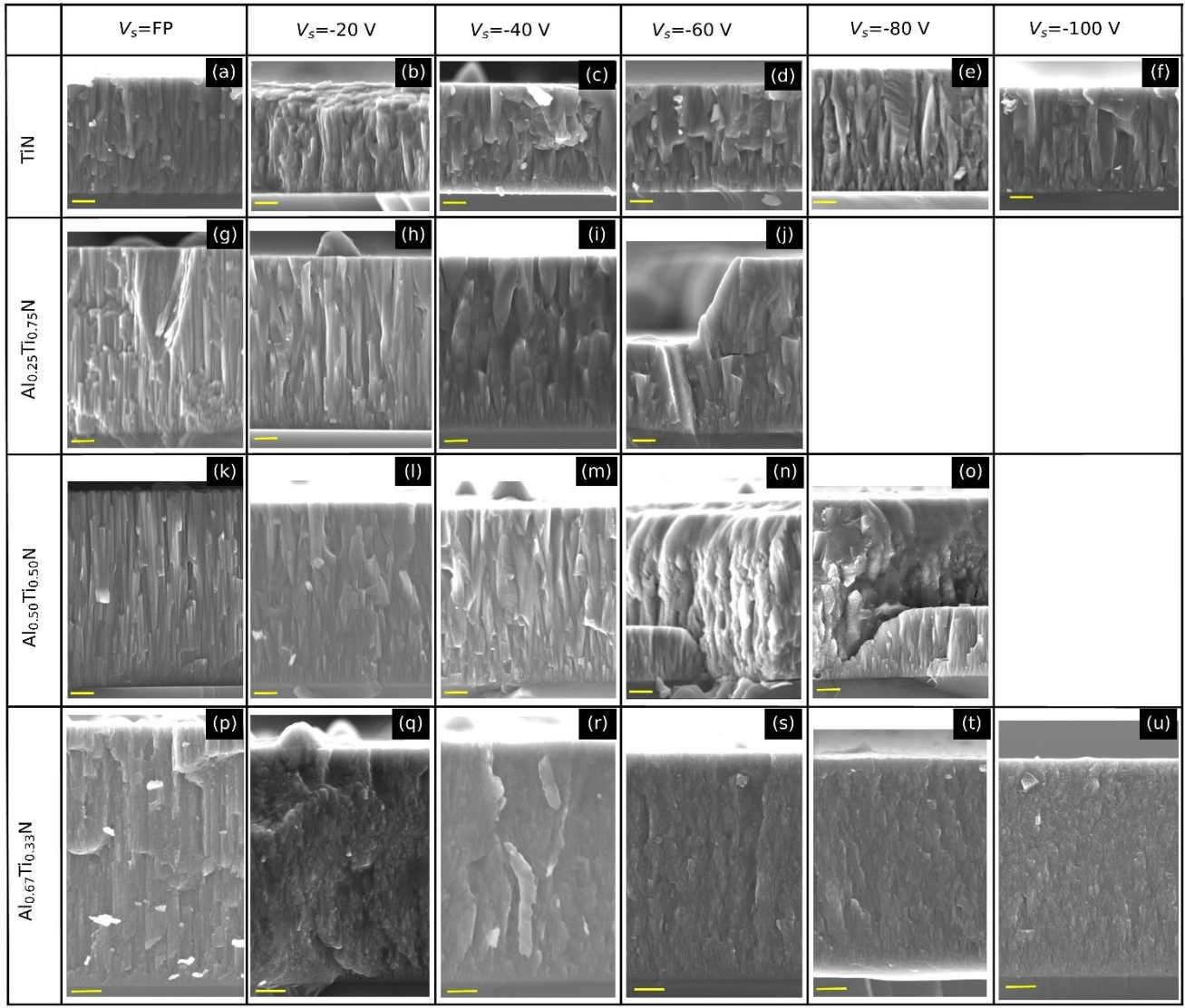

*Scale bar 200 nm

*Figure 3 shows fractured cross-sectional FESEM micrographs of* $Al_xTi_{1-x}N$ *coatings deposited on Si (100) substrate at different substrate bias voltage ($V_s$). The scale bar is 200 nm.*

Planview SEM micrographs of the TiN coatings grown at $V_s$ = FP and -20 V reveal a platelet-like surface morphology with the presence of nanometre size pores in coating grown with FP substrate bias (see dotted red circles in Figure S-2 of SI). TiN coatings grown with $|V_s| \geq 40$ V shows granular morphology indicating the formation of faceted nanocolumns. Surface morphologies of for $Al_{0.25}Ti_{0.25}N$ coatings deposited at $V_s$ = FP and -20 V are similar to the TiN coatings deposited at the same $V_s$. But coatings deposited at $|V_s| \geq 40$ V are different from those of TiN, where the top of nanocolumns are faceted but relatively flatter than those of TiN. Plan view SEM of $Al_{0.50}Ti_{0.50}N$ coating grown with $V_s$ = FP shows a clear view of sharp pyramidal faceted nanocolumns with the clear presence of voids between them. At higher $V_s$, the surface morphologies are compact and flat faceted. Unlike $Al_xTi_{1-x}N$: x ≤0.5, surface morphologies of $Al_{0.67}Ti_{0.33}N$ are granular and with the increase in bias the domain sizes of coatings are finer. Low magnification SEM images show cohesive failure and in-plane spallation of coatings (see Figure S-3 of SI).

To get a deeper insight into the microstructure, we probed the microstructure of TiN coatings ($V_s$ = FP and -40 V) with ex-situ TEM (see Figure 4). Clearly, below $t_f$ of 400 nm the domain sizes are different from those above it (see Figure 4**Error! Reference source not found.** (a-1 and b-1)). Towards the interfaces of coatings and substrate the domain sizes are smaller, and their number densities are higher compared to the top of the coatings. With $t_f$ larger than 400 nm, the in-plane domain sizes of the TiN coating deposited with $V_s$ = FP is almost constant at ~ 100 nm (see Figure 4 (a-1)). A closer look at the top of the coatings reveal that a few nanocolumns are separated from each other by a few nanometres size gaps (see arrow marked region of Figure 4 (a-2)). The selected area electron diffraction (SAED) pattern from overall region of films and substrate reveal a spotty diffractogram suggesting formation of crystalline and textured grains (see Figure 4 (a-3)). The predominant c-111 diffraction along the GD agrees with our in-situ XRD, suggesting no change in the grain orientation during the cooling down of the coatings as well as post-deposition. HR-TEM images recorded close to the interface of coating and substrate revealed that crystallites formed at $t_f$ close to ≈ 2 nm are having an interplanar spacing close to 0.21 nm, and hence crystal planes are identified as c-200 (see Figure 4 (a-4)). The c-200 crystal planes are oriented at an angle close to ≈ 55° to the direction normal to coating/substrate interfaces (see Figure 4 (a-4)). Comparing the 2D x-ray diffractograms recorded at the end of deposition (see Figure 2 (a)) along with the HRTEM image (see Figure 4 (a-4)) suggests, that from a $t_f$ of ≈ 2 nm, c-200 planes of TiN deposited with $V_s$ = FP remains at an angle close to 55° to the GD. An intracolumnar grain boundary is presented, see Figure 4 (a-5), indicating a distinct (see Figure 4 (a-4)) c-200 grains oriented in two different ψ-directions. The red triangle indicates an area of the sample which has an amorphous-like structure. Unlike TiN deposited at $V_s$ = FP, the intercolumnar gap vanished in coatings deposited with $V_s$ = -40 V (see arrow marked region in Figure 4 (b-2)). We also noticed that at $t_f$ lower than 400 nm, a large density of nanocolumn's growth is terminated (see arrow marked columns in Figure 4 (b-1)), beyond which a dense coating is seen. SAED patterns (see Figure 4 (b-3)) revealed a diffractogram corresponding to the random orientation of crystal structure, corroborating our XRD pattern (see Figure 2 (c)). HRTEM image from the coating/substrate interfaces region further revealed a poor crystalline quality of the grains and their random orientation exist from the initial stages of the deposition onwards. Furthermore, we probed the intracolumnar grain boundary (see arrow marked in Figure 4 (b-5)), which suggests that the grain boundaries are sharp and crystalline.

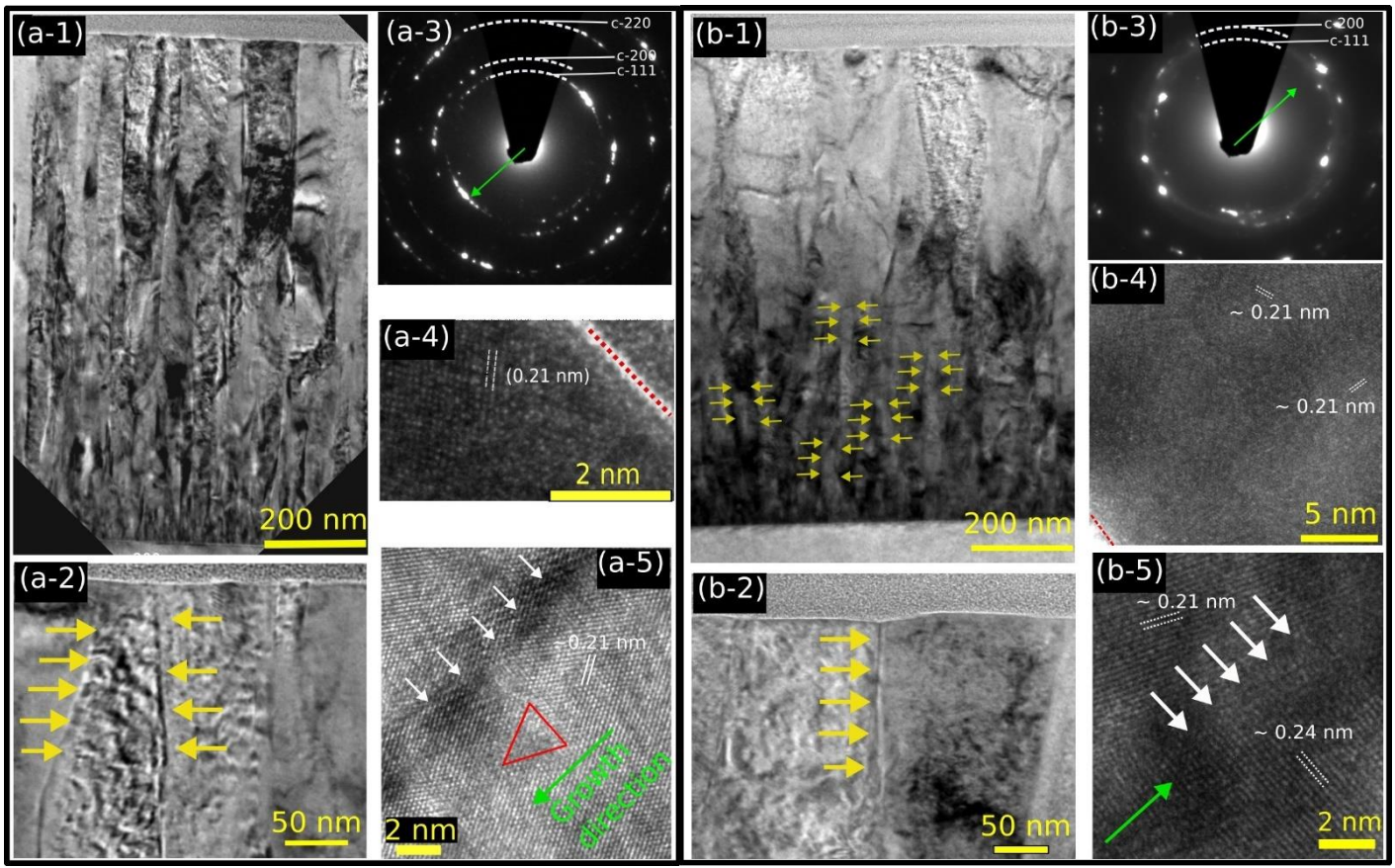

*Figure 4 is the TEM analysis of TiN coatings deposited at substrate bias voltage ($V_s$) of FP (a-1 to a-4), and -40 V (b-1 to b-4). Figure a-1 and a-2 are TEM images, a-3 is the SAED pattern, a-4, is the HRTEM image recorded at the interfaces of the coating and Si(100) substrate and a-5 a HRTEM image from the intracolumnar grain boundary. Figure b-1 and b-2 are TEM images of TiN coating deposited at a substrate bias voltage ($V_s$) of -40 V. Figure b-3 is the SAED pattern, b-4 is an HRTEM image of the interface, and b-5 is the intracolumnar grain boundary. The green coloured arrows depict the growth direction. Diffraction signal from the cubic phase is indexed in the SAED patterns. The red-coloured dashed line in b-4 represents the line of interfaces.*

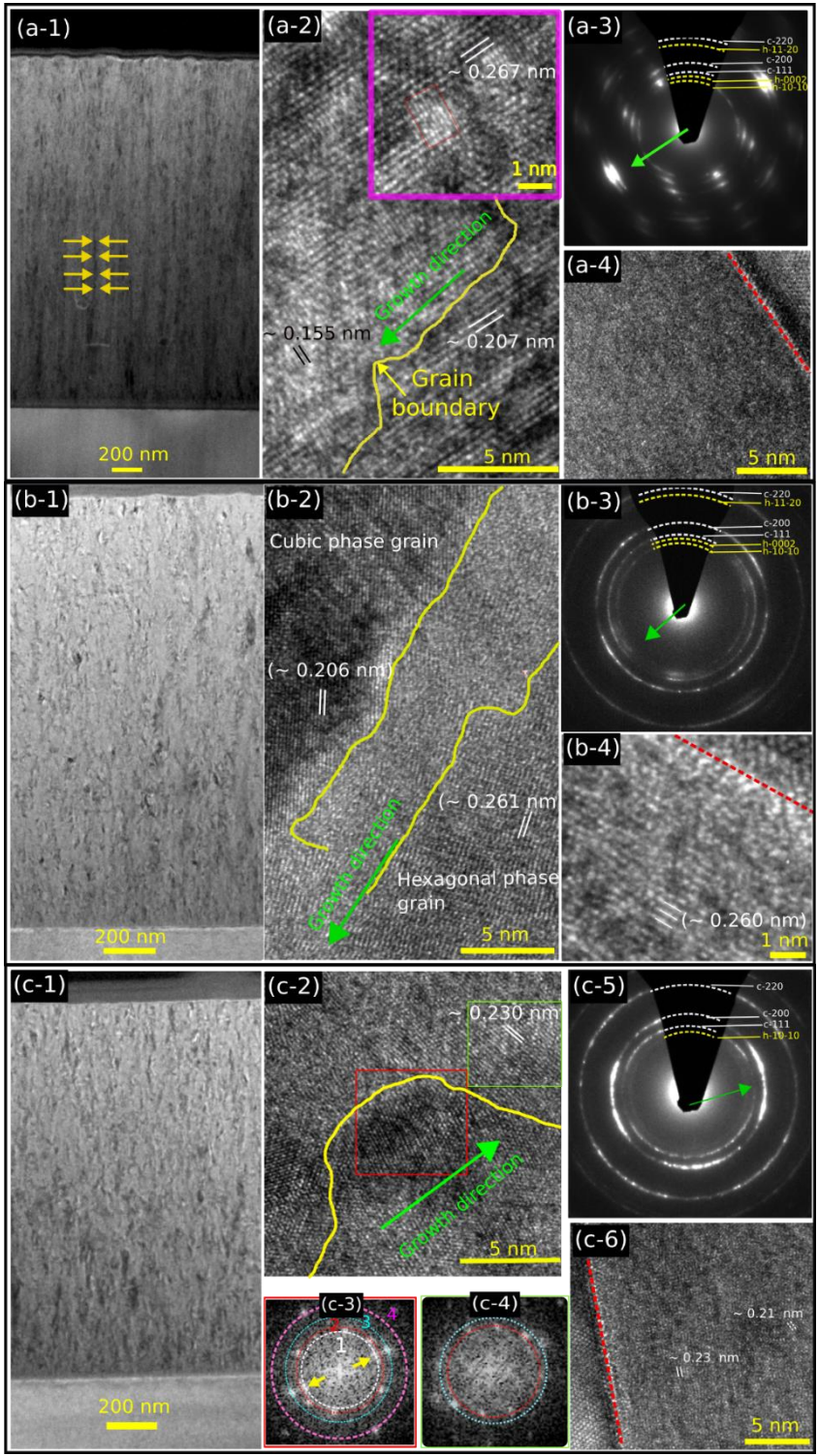

*Figure 5 the TEM analysis of $Al_{0.67}Ti_{0.33}N$ coatings deposited at substrate bias voltage ($V_s$) of FP (a-1 to a-4), -40 V (b-1 to b-4), and -80 V (c-1 to c-6). Figure a-1 is BF-TEM image, a-2 is an HRTEM image, a-3 the SAED pattern, and a-4 an HRTEM image recorded at the interface of the coating and Si(100) substrate. The inset of figure a-2 (area within the pink square) shows a cubic domain (within the red rectangle) separated and grain boundary, where formation of hexagonal phases are noticed. Figure b-1 is a BF-TEM image, b-2 a HRTEM image, b-3 is the SAED pattern, b-4 is an HRTEM image from the interface. Figure c-1 is the BF-TEM image and c-4 is an HRTEM image. Figure c-3 and c-4 are the FFT pattern of the region within the red square and green rectangle in c-2, respectively. The circle marked as 1, 2, 3, and 4 in figure c-3 are corresponding FFT patterns from h-10$\bar{1}$0, c-111, c-200, and c-220 crystal planes, respectively. and c-5 is the SAED pattern. c-6 is an HRTEM image of the interface. The green coloured arrow depicts the growth direction. The red coloured dashed line represents the interface between coating and substrate.*

Ex-situ microstructure of $Al_{0.67}Ti_{0.33}N$ coatings [$V_s$ = FP, -40 V, and -80 V] are further studied using TEM (see Figure 5). A *nanocrystalline* -like morphology is confirmed for $Al_{0.67}Ti_{0.33}N$ coatings (see Figure 5 a-1), b-1), and c-1)). While $Al_{0.67}Ti_{0.33}N$ coatings deposited with $V_s$ = -40 and -80 V are dense (see Figure 5 (b-1, and c-1)), an intercolumnar space is quite noticeable in coating deposited with $V_s$ = FP (see Figure 5 (a-1)). The presence of both hexagonal and cubic phases is further confirmed by both HRTEM and SAED. $Al_{0.67}Ti_{0.33}N$ coatings deposited with $V_s$ = FP and -40 V show the formation of hexagonal phases (see Figure 5 (a-2, a-3, b-2, b-3)). The formation of very small hexagonal and cubic domains is seen in $Al_{0.67}Ti_{0.33}N$ coating deposited with $V_s$ = FP (see inset of Figure 5 (a-2)). The interplanar spacing of 0.267 nm is corresponding to h-0002 of $Al_{0.67}Ti_{0.33}N$, which is aligned along IP direction (see inset of Figure 5 (a-2)) in agreement with our XRD (see Figure S-6 (a) of SI). The region marked with a red rectangle is corresponding to the cubic domain with c-220 ($d_{220}$ = 0.155 nm) crystal planes aligned along the GD (see inset of Figure 5 a-2)). . We also find that a distinct grain boundary between cubic and hexagonal phases exists along with a grain boundary between differently oriented cubic phases of $Al_{0.67}Ti_{0.33}N$ (see Figure 5 (a-2)). SAED patterns along with the HRTEM, of $Al_{0.67}Ti_{0.33}N$ coating deposited with $V_s$=FP, suggest that the h-$10\bar{1}0$, h-$11\bar{2}0$ and c-220 crystal planes are also oriented along the GD. This is in contrasts with the crystal orientation of $Al_xTi_{1-x}N$:x≤0.5, deposited at $V_s$=FP, coatings, where 111 crystallographic planes of cubic $Al_xTi_{1-x}N$ are oriented along the GD. Based on this observations we conclude that the orientation of crystallographic orientation in $Al_xTi_{1-x}N$ alloys in polycrystalline thin film form is not only constrained by geometry of the deposition system but can be influenced by the composition of the coatings too, especially in the compositions where mixed phases of crystals can form. Further inspection of HRTEM images of $Al_{0.67}Ti_{0.33}N$ coating with $V_s$ = FP, near coating/substrate interface regions, suggests the presence of an amorphous-like region (see Figure 5 a-4)).

Similar to $Al_{0.67}Ti_{0.33}N$ coating deposited with $V_s$ = FP, for coating deposited with $V_s$= -40 V, grain boundary between hexagonal and cubic phases are also seen (see Figure 5 (b-2)). However, the region in between the grain boundaries (as depicted between yellow lines in Figure 5 (b-2)) is either an amorphous-like region or crystalline in which zone-axis is not aligned with the beam directions. Nevertheless, a clear presence of grain boundaries is noticed. The recorded SAED pattern for coating with $V_s$ = -40 V is a bit different than that of the coating with $V_s$ = FP. The diffraction intensities from hexagonal phases are faint compared to cubic phases (see Figure 5 b-3)), further corroborating our XRD findings (see Figure 2(p)). The dominant crystallographic orientations are no longer c-220 and h-$11\bar{2}0$, but rather c-200 along the GD. The faint h-$11\bar{2}0$ pattern appears along the in-plane (see Figure 5 (b-3)). HRTEM images recorded close to the interface of coating/substrate revealed the formation of small crystallites with an interplanar spacing of ≈ 0.260 nm corresponding to interplanar spacing of h-0002 crystal planes of $Al_{0.67}Ti_{0.33}N$, and they are oriented along the GD, further corroborating our XRD findings.

In $Al_{0.67}Ti_{0.33}N$ coating deposited at $V_s$ = -80 V, our XRD data show a complete absence of hexagonal phases. However, the ex-situ SAED pattern revealed the presence of a very faint h-$10\bar{1}0$ pattern with no preferential orientations (see Figure 5 (c-5)). No detectable h-0002 or h-$11\bar{2}0$ signal is seen. A careful investigation of HRTEM images reveal that hexagonal phases are formed at the grain boundary region only. The fast-Fourier transformation (FFT) of the HRTEM image around a grain boundary (see the red squared area in Figure 5 c-2)) shows the presence of h-$10\bar{1}0$ spots (see spots aligned with circle denoted as 1 in Figure 5 c-3)). The FFT pattern (see Figure 5 c-4)) obtained from an individual grain (green squared area of Figure 5 (c-4)) revealed a complete absence of any spots corresponding to the hexagonal phases. Thus, we conclude that for $Al_{0.67}Ti_{0.33}N$ coatings deposited at higher $V_s$ [e.g., -80 and -100 V], the hexagonal phase nucleates only at grain boundaries, but not in the bulk of the grains. Further, an HRTEM image taken at the interface between coatings and substrate revealed the formation of very small crystallites with c-200 crystal planes oriented along the GD, and c-111 inclined with an angle of close to 45° (see Figure 5 c-6)) [≈44° measured from XRD]. This observation establishes that the crystal planes of the coatings seen at very small film thickness (~2-3 nm) retain their crystallographic orientation throughout the deposition.

To further understand the temporal evolution (i.e., average grain size as a function of film thickness increases) of the above-discussed grain (also crystallite) sizes and their orientation (i.e., texture) during the thin film depositions, we analysed the line profiles and diffraction intensities of the recorded diffractograms. While the thickness-dependent evolution of x-ray diffractograms and their evolution along the GD is thoroughly discussed in the section SI. III of SI, we dedicate the next section to the evolution of the grain/crystallite sizes.

### C. Temporal evolution of crystallite size ($D_{hkl}$)

Many film properties such as, film hardness and stress evolution strongly connected to the grain size of the coatings. Thus, here we evaluate the temporal evolutions of crystallite size ($\langle D_{hkl} \rangle$) of $Al_xTi_{1-x}N$ with different $V_s$ along the GD (see Figure S-9 of SI) as well as close to the IP direction (see Figure 6 (a-d)). For all compositions, the respective ranges of ψ values that are used for integrating the diffraction intensities to obtain the line profiles are mentioned on top of each figure (see Figure 6). While here we discuss the crystallite size evolution along the *in-plane* direction, details of the same along GDs are discussed in section SI. IV of SI.

We find that a power-law like (see Eqn. 2) increase in average $D_{111}$ as film thickness increases is noticed for x= [0,0.5] in $Al_xTi_{1-x}N$. Thin films of $Al_xTi_{1-x}N$:x= [0,0.5] deposited with $V_s$=FP, shows a rapid increase in $D_{111}$ as $t_f$ increases as compared to films deposited with a finite applied $V_s$ (see Figure 6(a-c)). For example, at $t_f$ ≈ 200 nm, $D_{111}^{IP}$ of TiN deposited at $V_s$ = FP is estimated to ≈ 23 nm and increased to ≈ 38 nm at $t_f$ =1007 nm while film deposited at $V_s$ =-20 V at the end of depositions ($t_f$ ≈ 920 nm) $D_{111}^{IP}$ is estimated to ≈ 37 nm. With further increase in $V_s$ to -40, -60, -80, and -100 V, $D_{111}^{IP}$ of TiN at the end of depositions are ≈ 29 nm, ≈ 30 nm, ≈ 33 nm, and ≈ 33 nm, respectively (see Figure 6 (a)). Similar behaviour on the evolution of $D_{111}^{IP}$ in $Al_{0.25}Ti_{0.75}N$ and $Al_{0.50}Ti_{0.50}N$ coatings is seen. Thin film of $Al_{0.25}Ti_{0.75}N$ deposited with $V_s$ = FP increases from ≈19 nm [$t_f$ ≈ 200 nm] to ≈41 nm [$t_f$ ≈ 1631 nm] (see Figure 6 (b)). With $V_s$ = -20 V, $D_{111}^{IP}$ is estimated to ≈ 23 nm at $t_f$ ≈ 200 nm and increases to ≈33 nm as $t_f$ reaches to 1494 nm (see Figure 6 (b)). A similar evolution of $D_{111}^{IP}$ is observed for $V_s$ = -40 V. At $t_f \approx 200$ nm, $D_{111}^{IP}$ is estimated to $\approx 22$ nm, while as $t_f$ reaches to 1494 nm, $D_{111}^{IP}$ is $\approx 33$ nm. Likewise, the texture evolution of $Al_{0.25}Ti_{0.75}N$ at $V_s$=-60 V (see Figure S-9(b) of SI), a discontinuity in evolution of $D_{111}^{IP}$ is also observed at $t_f \approx 770$ nm (see Figure 6 (b)). The $D_{111}^{IP}$ at $t_f \approx 200$ nm is $\approx 24$ nm, while at $t_f \approx 770$ nm it is estimated to be $\approx 27$ nm in this case. Similar to $Al_{0.25}Ti_{0.75}N$ coatings the evolutions of $D_{111}^{IP}$ vs. $t_f$ in $Al_{0.50}Ti_{0.50}N$ coatings deposited with $V_s$= -60 and -80 V show an abrupt change at $t_f \approx 950$ nm and 745 nm, respectively (see Figure 6 (c)). At $t_f \approx 150$ nm, $D_{111}^{IP}$ of $Al_{0.50}Ti_{0.50}N$ coating deposited at $V_s$ = -60 is estimated to $\approx 18$ nm, while at $t_f \approx 950$ nm $D_{111}^{IP}$ is 25 nm. $D_{111}^{IP}$ of $Al_{0.50}Ti_{0.50}N$ coating deposited at $V_s$ = -80 V changes from 16 nm to 23 nm, as $t_f$ increases from $\approx 150$ nm to $\approx 1545$ nm.

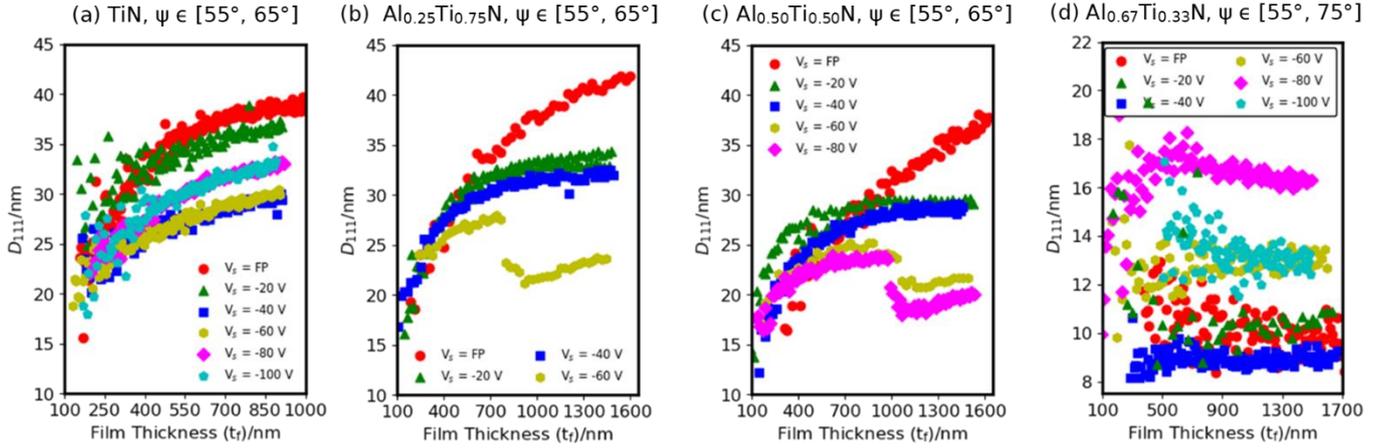

*Figure 6 (a-d) are the temporal evolution of crystallite size ($D_{hkl}$) of $Al_xTi_{1-x}N$ close to in-plane direction. Thin film coatings are deposited on Si (100) substrate at various substrate bias voltages. The range of tilt angle($\psi$) used to integrate the intensity of XRD patterns in transforming 2D XRD to 1D line-out are shown on top of each figure.*

The $D_{111}^{IP}$ of $Al_{0.67}Ti_{0.33}N$ coatings are estimated by integrating the diffraction intensities from c-111 crystal planes within the $\psi$ ranges of 55-75° (see Figure 6 (d)). $Al_{0.67}Ti_{0.33}N$ coating deposited with $V_s$ = FP, estimated $D_{111}^{IP}$ at $t_f \approx 400$ nm is $\approx 11$ nm, which reduced to $\approx 9.5$ nm at $t_f \approx 1867$ nm. Estimated $D_{111}^{IP}$ of $Al_{0.67}Ti_{0.33}N$ coating with $V_s$ = -20 V is 10.5 nm at all $t_f$. The evolution of $D_{111}^{IP}$ of $Al_{0.67}Ti_{0.33}N$ coating deposited with $V_s$ = -40 V is a bit complex. From $t_f \approx 280$ nm to $\approx 650$ nm, a small increase in $D_{111}^{IP}$ from 8.6 nm to 9.6 nm is noticed. Beyond $t_f \approx 650$ nm, $D_{111}^{IP}$ decreases to 8.6 nm and almost remains constant thereafter. With $V_s$ = -60 V, $D_{111}^{IP}$ remains almost constant at 12.9 nm. The average $D_{111}^{IP}$ of $Al_{0.67}Ti_{0.33}N$ coating with $V_s$ = -60 V is again a bit complex. We observe between the $t_f \approx 120$-550 nm that $D_{111}^{IP}$ increases from 11.4 nm to 17.5 nm. Beyond $t_f \approx 550$ nm, $D_{111}^{IP}$ decreases with an increase in $t_f$, and at $t_f \approx 1508$ nm $D_{111}^{IP}$ is estimated to be 16.5 nm. The estimated $D_{111}^{IP}$ of $Al_{0.67}Ti_{0.33}N$ coating deposited with $V_s$ = -100 V is 14.5 nm at $t_f \approx 500$ nm, and it decreases to 13 nm at $t_f \approx 1502$ nm. The evolutions of $D_{111}^{IP}$ vs. $t_f$ for all samples are fitted with Eqn. 2, and the coefficients (***k***) and exponents (***n***) are tabulated in Table S-2 of SI.

Next, to uncover the effect of compositions and substrate bias (and consequently the formation and evolution of different microstructures as discussed in previous sections) to the stress evolutions of the films, the temporal evolution of the thickness averaged stress is determined (see section II and section SI. I of SI for details of evaluation method). As mentioned in section II, the film thickness, $t_f$, averaged stress is modelled with different two numerical equations. Below we presented evolution of internal stress (determined from internal strain, see Method section) in TiN, $Al_{0.25}Ti_{0.75}N$, $Al_{0.50}Ti_{0.50}N$, and $Al_{0.67}Ti_{0.33}N$ in four different sub-sections of section IV.

## IV. EVOLUTION OF BIAXIAL STRAIN AND STRESS

### A. Titanium nitride (TiN)

We determined both biaxial IP strain ($\varepsilon_{\psi=90°}$) and GD strain ($\varepsilon_{\psi=0°}$) using the diffractograms from c-111 crystal planes of all coatings and monitored their temporal evolution (see Figure S-11 of SI). The biaxial stress ($\sigma$) is shown in Figure 7. The negative sign (-) in the strain/stress values indicate all TiN films are experiencing a compressive strain/stress. The positive (+) sign during thin film deposition represents tensile strain/stress. In these experiments, we could assess stress/strain from diffractograms recorded for $t_f \geq 200$ nm only, due to too low intensity at lower thicknesses.

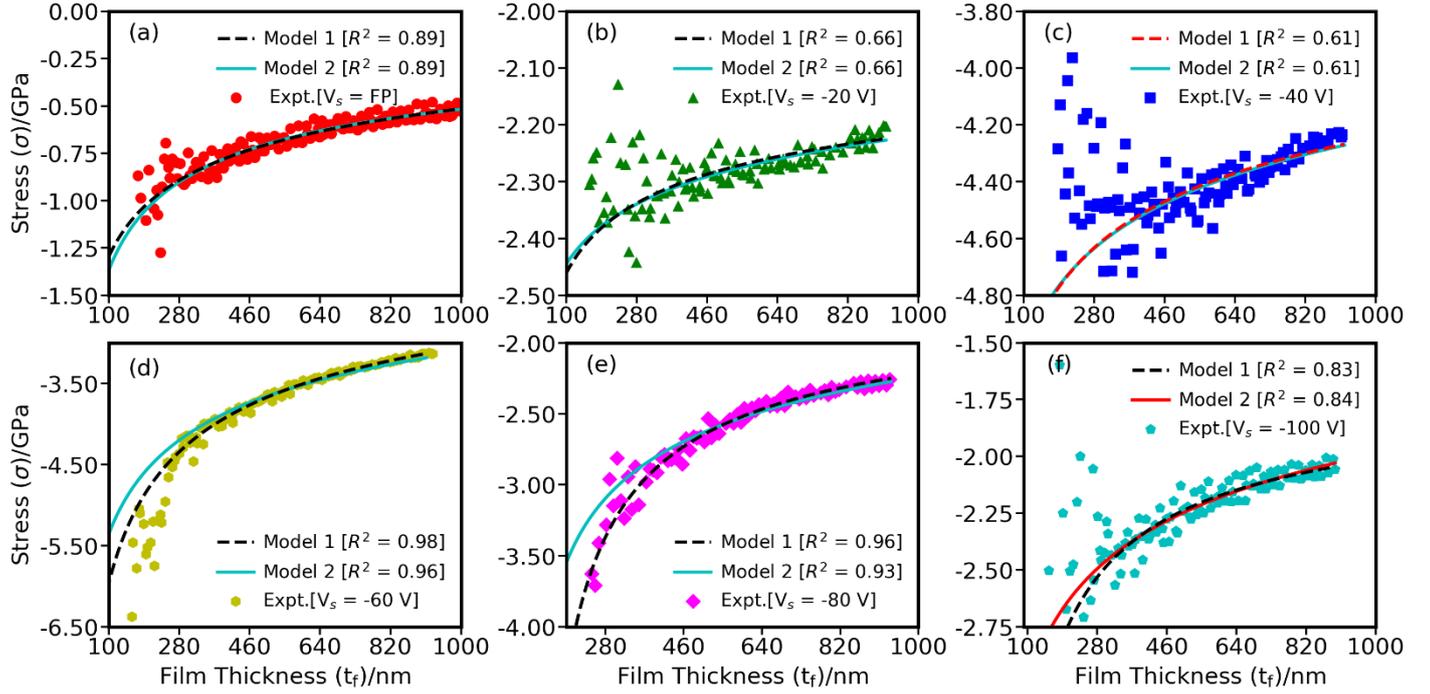

*Figure 7 is the biaxial stress ($\langle\sigma_{111}\rangle$) evolution of TiN on Si (100) substrate at different substrate bias voltages $V_s$; at floating potential (a), -20 V (b), -40 V (c), -60 V (d), -80 V (e), and -100 V (f). Experimentally obtained data are fitted with a power law (Model 1, dashed line) and a kinetic model (Model 2, solid lines) as discussed in section II.*

For TiN, deposited at all $V_s$, the final $\sigma$ is always compressive. At $t_f \approx 280$ nm, $\sigma$ of TiN coatings deposited at $V_s$ = FP, -20, -40, -60, -80, and -100 V are approximately -0.8, -2.3, -4.5, -4.3, -3.0, and -2.5 GPa, respectively. At the very end of the deposition, the $\sigma$ (and $t_f$) is estimated as -0.5 GPa ($\approx$ 1007 nm), -2.2 GPa ($\approx$ 917 nm), -4.2 GPa ($\approx$ 920 nm), -3.1 GPa ($\approx$ 912 nm), -2.2 GPa ($\approx$ 930 nm), and -2.0 GPa ($\approx$ 895 nm) for $V_s$ = FP, -20, -40, -60, -80, and -100 V, respectively. This suggests a "U" type behaviour of final values of $\sigma$ in TiN films with increase in $V_s$. We use the power law (presented in Eqn. 3) to fit the $\sigma$ vs. $t_f$ (see dashed lines (Model 1) in Figure 7). The fitted co-efficient and exponents for TiN coatings with varying $V_s$ are tabulated Table 2. It shows that, the parameter $\sigma_0$, to which $\sigma(t_f)$ converge was found to be positive (0.54 GPa) for coating deposited with $V_s$ = FP, while for coatings deposited with $V_s$ of -20, -40, -60, -80 and -100 V, the $\sigma_0$ is negative but have a U type dependence to the $V_s$ (see Table 2). The coefficient c, which is related to the stress generated or shrinkage by grain boundaries, [55,57] is always a negative number. The values of $\gamma$ are close to 0.20 with $V_s$ = FP, -20, and -40V and consistently increases with further increases in $V_s$ (see Table 2). We note that for TiN coatings deposited at $V_s$ = -20 and -40 V, the estimated stress in between 280-460 nm is a bit dispersed, suggesting a wider distribution of stress present in the coatings at such lower $t_f$.

*Table 2 Least squares optimized pre-factors and exponents obtained by fitting the evolution of stress ($\langle\sigma_{111}\rangle$) vs. $t_f$ in $Al_xTi_{1-x}N$ with Eqn. 3 (Model 1).*

| $V_s$ (V) | $\sigma_0$ (GPa) | | | | c (GPa nm$^\gamma$) | | | | $\gamma$ | | | |
|---|---|---|---|---|---|---|---|---|---|---|---|---|
| | x=0.0 | x=0.25 | x=0.50 | x=0.67 | x=0.0 | x=0.25 | x=0.50 | x=0.67 | x=0.0 | x=0.25 | x=0.50 | x=0.67 |
| FP | 0.54 | 0.52 | 0.83 | 1.1 | -5.54 | -3.47 | -8.93 | 3474.1 | 0.24 | 0.2 | 0.37 | 1.29 |
| -20 | -1.80 | -0.77 | -0.94 | -1.3 | -1.66 | -2.19 | -6.18 | 3474.1 | 0.20 | 0.2 | 0.46 | 1.35 |
| -40 | -2.91 | -2.24 | -2.01 | -3.2 | -5.32 | -2.97 | -16.38 | 3474.1 | 0.20 | 0.2 | 0.56 | 1.35 |
| -60 | -1.03 | -4.28 | -3.10 | -4.3 | -29.07 | -2.52 | -84.70 | 3474.1 | 0.38 | 0.2 | 0.60 | 1.58 |
| -80 | -1.56 | Ø | Ø | -5.8 | -49.90 | Ø | Ø | -6.1 | 0.63 | Ø | Ø | 0.25 |
| -100 | -1.66 | Ø | Ø | Ø | -45.08 | Ø | Ø | Ø | 0.70 | Ø | Ø | Ø |

The σ vs. $t_f$ of TiN coatings are further analysed within the kinetic model proposed by Chason *et al.* [56], which is briefly discussed in section II. The $\sigma_c$ and $\sigma_{T,0}$ (L =100 nm) are fixed at -0.02 and 0.67 GPa, respectively. The optimized coefficients and exponents obtained by least-squares fitting of Eqn. 4 (shown as solid lines (Model 2) in Figure 7) with the experimental data are tabulated in The value of $\beta D_{eff}$, which measures the effective diffusivity of adatoms, is estimated to 47.6 nm$^2$/s at $V_s$ = FP. With applied substrate bias [$V_s$ = -20 V to -40 V], $\beta D_{eff}$ increases by a factor of at least two compared to FP. TiN films deposited with $V_s$= -60 to -100 V, $\beta D_{eff}$ is smaller compared to lower $V_s$, but slightly increases with increase in $|V_s|$. The value of $l$, which measures the surface to the depth of defect generated by the energetics particle bombardments as well as the region of grain boundary densification increases from 0.21 nm (at $V_s$ = FP) to ~ 0.7 nm (at $V_s$ =-100 V). The least-squares fitted value of A,B, and $D_i$ is tabulated in Table 3, in which, $D_i$ is the diffusivity of defects created in the bulk of coatings that diffused to the surface. The best fitted values of $D_i$ have a non-linear dependence to $V_s$ and are 72.9, 23.7, 37.3, 75.1, 74.8, and 59.5 nm$^2$/s at $V_s$ = FP, -20, -40, -60, -80, and -100 V, respectively.

*Table 3 Least squares optimized pre-factors and exponents obtained by fitting the evolution of stress ($\langle\sigma_{111}\rangle$) vs. $t_f$ in Al$_x$Ti$_{1-x}$N with Eqn. 4 (Model 2).*

| $V_s$ (V) | $\beta D_{eff}$ (nm$^2$/s) | | | | $l$ (nm) | | | | A (GPa) | | | | B (GPa) | | | | $D_i$ (nm$^2$/s) | | | |
|---|---|---|---|---|---|---|---|---|---|---|---|---|---|---|---|---|---|---|---|---|
| | x=0.0 | x=0.25 | x=0.50 | x=0.67 | x=0.0 | x=0.25 | x=0.50 | x=0.67 | x=0.0 | x=0.25 | x=0.50 | x=0.67 | x=0.0 | x=0.25 | x=0.50 | x=0.67 | x=0.0 | x=0.25 | x=0.50 | x=0.67 |
| FP | 47.6 | 81.4 | 87.5 | 50.1 | 0.21 | 0.43 | 0.30 | 0.30 | -122.4 | -25.9 | -27.2 | -12.0 | -82.1 | -27.8 | -20.7 | 0.0 | 72.9 | 49.9 | 92.3 | Ø |
| -20 | 143.9 | 201.0 | 147.6 | 70.2 | 0.30 | 0.46 | 0.44 | 0.56 | -50.5 | -48.8 | -49.2 | -31.6 | -96.4 | -35.0 | -29.9 | -20.1 | 23.7 | 57.5 | 49.7 | 89.9 |
| -40 | 111.3 | 209.9 | 240.1 | 99.7 | 0.56 | 0.75 | 0.56 | 0.70 | -82.9 | -48.3 | -60.7 | -37.7 | -118.1 | -67.6 | -49.5 | -29.4 | 37.3 | 68.1 | 52.0 | 74.8 |
| -60 | 60.0 | 76.6 | 42.6 | 194.75 | 0.76 | 0.80 | -0.68 | 0.80 | -149.9 | -43.1 | -149.8 | -51.5 | -45.0 | -145.7 | -65.1 | -78.5 | 75.1 | 68.4 | 75.0 | 87.6 |
| -80 | 66.6 | Ø | Ø | Ø | 0.50 | Ø | Ø | Ø | -134.9 | Ø | Ø | Ø | -50.5 | Ø | Ø | Ø | 74.8 | Ø | Ø | Ø |
| -100 | 75.4 | Ø | Ø | Ø | 0.67 | Ø | Ø | Ø | -68.2 | Ø | Ø | Ø | -47.2 | Ø | Ø | Ø | 59.5 | Ø | Ø | Ø |

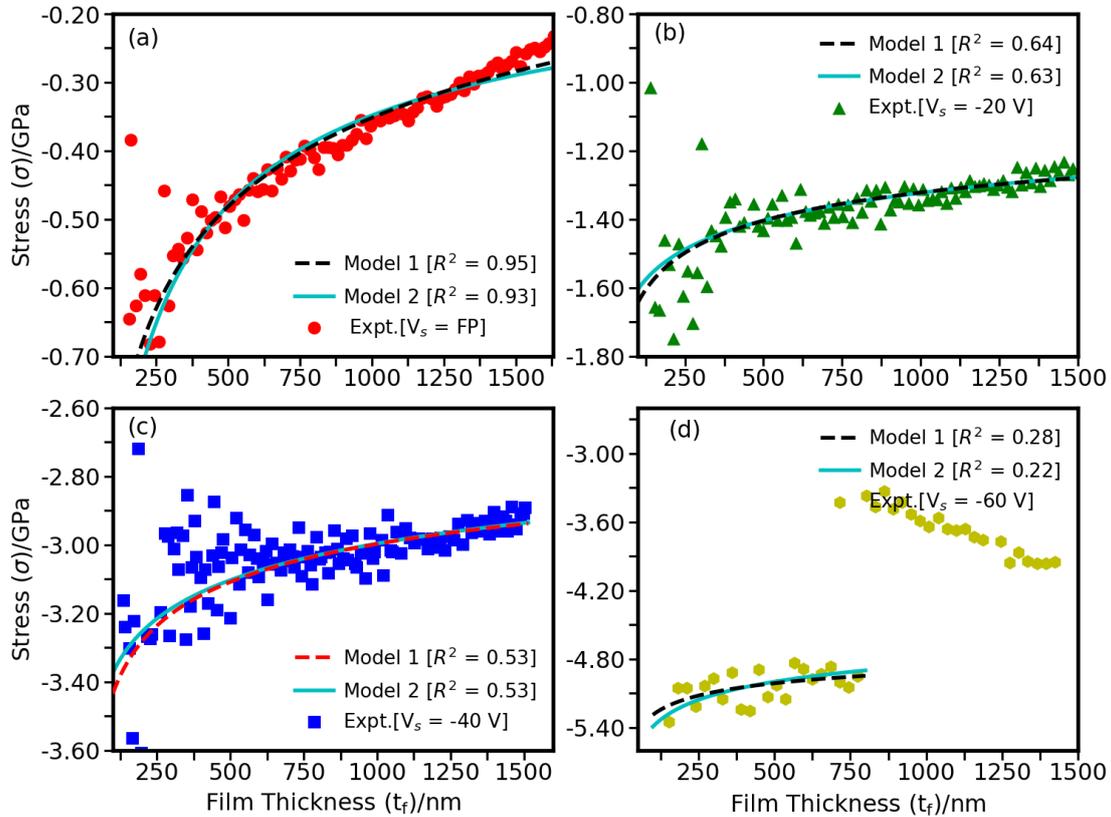

**B. Aluminium (25 at. %) titanium nitride (Al$_{0.25}$Ti$_{0.75}$N)**

*Figure 8 is the biaxial stress ($\langle\sigma_{111}\rangle$) evolution of $Al_{0.25}Ti_{0.75}N$ on Si (100) substrate at different substrate bias voltages $V_s$; at floating potential (a), -20 V (b), -40 V (c), and -60 V (d). Experimentally obtained data are fitted with a power law (Model-1, dashed line) and a kinetic model (Model-2, solid lines) as discussed in section II.*

The evolutions of $\sigma$ vs $t_f$ in $Al_{0.25}Ti_{0.75}N$ coatings at different $V_s$ [FP, -20, -40, and -60 V] are shown in Figure 8. The IP-strain and IP interplanar spacing of $Al_{0.25}Ti_{0.75}N$ as a function of $t_f$ are provided in Figure S-12 of SI. At $t_f \approx 250$ nm, the $\sigma$ is approximately -0.7, -1.6, -3.2, and -5.1 GPa for coatings deposited with $V_s$ = FP, -20, -40, and -60 V, respectively. For coatings deposited with the $V_s$ = FP, -20, and -40 V, $\sigma$ reduces with the increase in $t_f$. At the end of their deposition, the $\sigma$ is estimated to -0.2, -1.2, and -2.9 GPa at $V_s$ =FP, -20, and -40 V, respectively, which suggests the increase in compressive stresses in the coating with -ve substrate bias. Similar to TiN, $Al_{0.25}Ti_{0.75}N$ coatings deposited with $V_s$ = -20 and -40 V, the value of $\sigma$ is bit widespread. Unlike a smooth $\sigma$ evolution of previously discussed TiN coatings, $Al_{0.25}Ti_{0.75}N$ with a $V_s$ of -60 V shows a discontinuity in it at $t_f \approx 770$ nm, and $\sigma$ changes from -4.94 GPa to -3.33 GPa (see Figure 8(d)). In the thickness region of $t_f \geq 770$ nm, $\sigma$ vs. $t_f$ is fitted with Eqn. 7, where $\sigma_0$, c and $\gamma$ are obtained as -4.28 GPa, -2.52 GPa nm$^\gamma$ and 0.2 respectively. The least-squared fitted value of $\sigma_0$ and c from Eqn. 3 are tabulated in Table 2. A consistent increase in absolute value of $\sigma_0$ with increase in $V_s$ is noticed (see Table 2). The $\gamma$ remain constant at 0.2 in all these three coatings (see Table 2).

Least-squares fitting of $\sigma$ vs. $t_f$ with Eqn. 4, with a fixed $\sigma_c$ = -0.02 GPa and $\sigma_{T,0}$ = 0.85 GPa at $L_0$=100 nm, revealed that at $V_s$ = FP, $\beta D_{eff}$ is 81.4 nm$^2$/s and increases more than two-fold upon substrate bias ($V_s$ = -20 and -40 V, see Table 3). Like TiN, in $Al_{0.25}Ti_{0.75}N$, the $l$ increases from 0.43 nm to 0.75 nm with an increase in $V_s$ from FP to -40 V (see Table 3). By varying $V_s$, the values of A changed from -25.9 to -43.1, while B increased from -27.8 GPa to -145.7 GPa, as $V_s$ increases from FP to -40 V (see Table 3). The $D_i$ also increases with an increase in $V_s$ and changes from 48.9 to 68.4 nm$^2$/s, as $V_s$ changes from FP to -40 V (see Table 3). As mentioned previously $Al_{0.25}Ti_{0.75}N$ coating deposited with $V_s$= -60 V shows an abrupt change in stress at $t_f \approx 770$ nm, we analysed the stress evolution of the coating prior to this discontinuity. Least-squared fitting of $\sigma$ vs. $t_f$ in the given region of the coating revealed a relatively smaller $\beta D_{eff}$ ($\approx$ 76.6 nm$^2$/s) than that of other samples deposited with $V_s$ of -20 and -40 V (see Table 3).

### C. Aluminium (50 at. %) titanium nitride ($Al_{0.50}Ti_{0.50}N$)

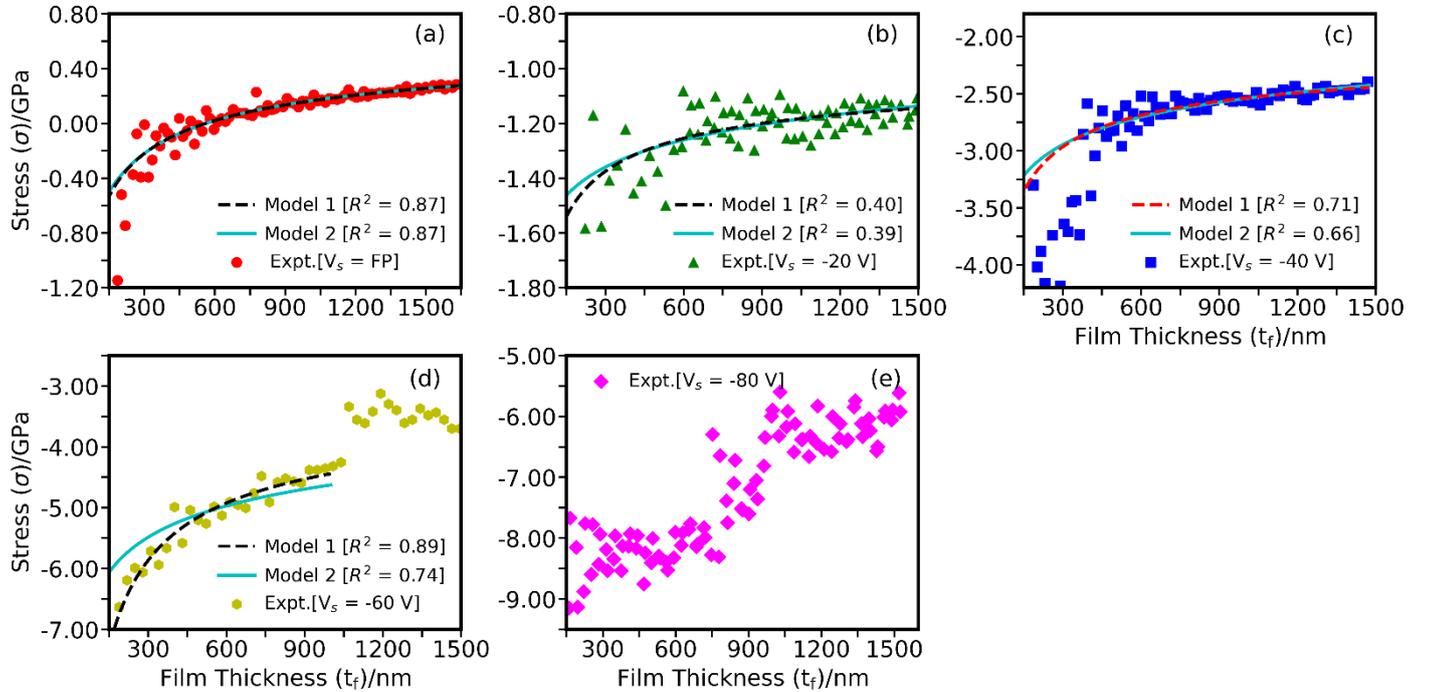

*Figure 9 is the biaxial stress ($\langle\sigma_{111}\rangle$) evolution of $Al_{0.50}Ti_{0.50}N$ on Si (100) substrate at different substrate bias voltages $V_s$; at floating potential (a), -20 V (b), -40 V (c), -60 V (d), and -80 V (e). Experimentally obtained data are fitted with a power law (Model 1, dashed line) and a kinetic model (Model 2, solid lines) as discussed in section II.*

Unlike TiN and $Al_{0.25}Ti_{0.75}N$, $Al_{0.50}Ti_{0.50}N$ coating deposited with $V_s$ = FP shows average tensile stress in the coating beyond $t_f \gtrsim$ 600 nm, while between $t_f \approx$ 185 nm to 600 nm $\sigma$ remains compressive (see Figure 9 (a)). At $t_f$ = 185 nm, the estimated $\sigma$ is -0.5 GPa, while $\sigma$ shows a tensile nature and it reaches to 0.3 GPa at $t_f$ = 1640 nm. With negative potential $V_s$, $\sigma$ is compressive throughout $t_f$. The $Al_{0.50}Ti_{0.50}N$ coating deposited at $V_s$ = -20 V shows a relatively weak evolution of $\sigma$ (see Figure 9 (b)). At $t_f \approx$ 250 nm, $\sigma$ is estimated to $\approx$ -1.4 GPa, while $\sigma$ converges to -1.1 GPa at $t_f \approx$ 1545 nm. For the coating deposited with $V_s$ = -40 V, we find a very strong stress relaxation between $t_f \approx$ 180-380 nm (see Figure 9 (c)). At $t_f \approx$ 180 nm, estimated $\sigma$ is -4.0 GPa, while at $t_f \approx$ 380 nm $\sigma$ is lowered down to -2.8 GPa. Similar to $Al_{0.25}Ti_{0.75}N$ coating deposited at $V_s$ = -60 V, there is a discontinuity in $\sigma$ vs. $t_f$ of $Al_{0.50}Ti_{0.50}N$ coating at the same $V_s$. At $t_f \approx$ 950-1000 nm, the $\sigma$ changes from -4.1 GPa to -3.3 GPa (see Figure 9 (d)). In this coating the estimated $\sigma$ at $t_f \approx$ 180 nm, is -6.5 GPa, while at $t_f \approx$ 950 nm it is -4.2 GPa. $Al_{0.50}Ti_{0.50}N$ coating deposited at $V_s$ = -80 V shows almost a constant $\sigma$ of -8.2 GPa for 170 nm $\lesssim t_f \lesssim$ 745 nm (see Figure 9 (e)). Beyond $t_f \gtrsim$ 745 nm, an abrupt change in $\sigma$ vs. $t_f$ is observed. The $\sigma$ evolution is fitted with Eqn. 3, and the coefficients and exponent are tabulated in Table 2. It is quite clear that the magnitude of converged stress $\sigma_0$, as well as $\gamma$, increases with an increase in $V_s$.

Fitting of the kinetic model, Eqn. 4, to gain the evolution of $\sigma$, with $\sigma_c$ = -0.02 GPa and $\sigma_{T,0}$ = 0.85 GPa at $L_0$=100 nm, revealed the two to three fold increase in $\beta D_{eff}$ value upon $V_s$= -20 and -40 V compared to $V_s$ = FP (see Table 3). Fitting of Eqn. 4 to $\sigma$ vs. $t_f$, 200 nm $\lesssim t_f \lesssim$ 920 nm, of $Al_{0.50}Ti_{0.50}N$ with $V_s$= -60 V, revealed $\beta D_{eff}$ to be much smaller than the coating deposited with $V_s$ = FP (see Table 3). The value of $l$ increases consistently from 0.30 nm to 0.68 nm with an increase in $V_s$ from FP to -60 V. The optimized A [and B] are tabulated in Table 3, where a consistent increase in magnitude of A and B is seen with increase in $|V_s|$. The best fitting of $D_i$ is obtained as 92.3, 49.7, 52.0, and 75.0 nm$^2$/s for $V_s$ = FP, -20 V, -40 V, and -60 V, respectively (see Table 3).

### D. Aluminium (67 at. %) titanium nitride ($Al_{0.67}Ti_{0.33}N$)

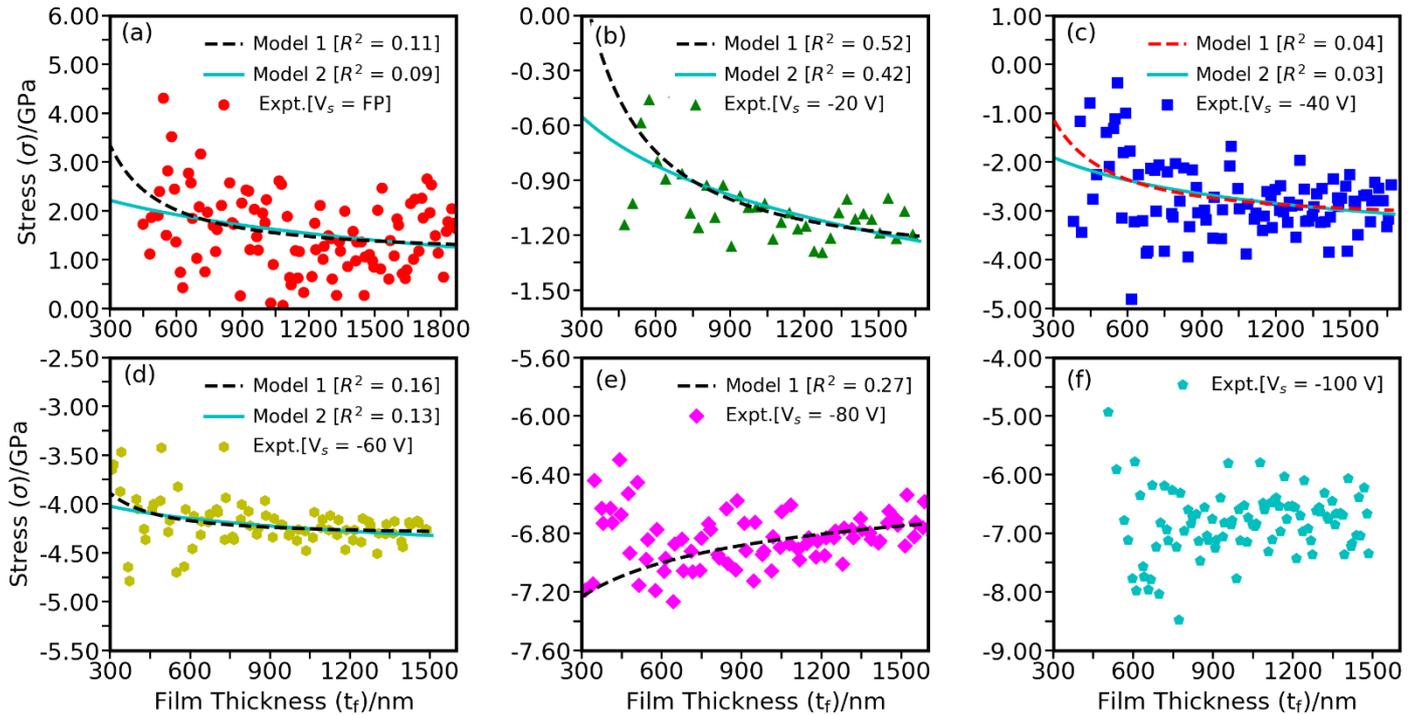

*Figure 10 is the biaxial stress ($\langle\sigma_{111}\rangle$) evolution of $Al_{0.67}Ti_{0.33}N$ on Si (100) substrate at different substrate bias voltages $V_s$; at floating potential (a), -20 V (b), -40 V (c), -60 V (d), -80 V (e), and -100 V (f). Experimentally obtained data are fitted with a power law (Model 1, dashed line) and a kinetic model (Model 2, solid lines) as discussed in section II.*

Unlike $Al_xTi_{1-x}N$: x $\leq$ 0.5, a wider distribution of $\sigma$ values are observed in $Al_{0.67}Ti_{0.33}N$ coatings (see Figure 10). The $Al_{0.67}Ti_{0.33}N$ coating deposited with $V_s$ = FP shows a net tensile stress with very high values of stress distribution within the recorded $t_f$ (300 nm $\lesssim t_f \lesssim$ 1870 nm) with a mean $\sigma$ of 1.6 GPa. With an increase in $t_f$, a reduction in the amplitude of tensile stress is noticeable (see Figure 10 (a)). The $Al_{0.67}Ti_{0.33}N$ coating with $V_s$ = -20 V a net compressive stress is estimated and the magnitude of compressive stress increases with an increase in $t_f$ (see Figure 10 (b)). A similar $\sigma$ evolution is observed in $Al_{0.67}Ti_{0.33}N$ coatings deposited with $V_s$ = -40 and -60 V (see Figure 10 (b-c)). The average $\sigma$ recorded with $Al_{0.67}Ti_{0.33}N$ coating with $V_s$= -40 V is ≈ -3.9 GPa. With $V_s$ = -60 V, the evolution of $\sigma$ is relatively better resolved than that of other $Al_{0.67}Ti_{0.33}N$ coatings (see Figure 10 (d)), and $\sigma$ increases with an increase in $t_f$. At $t_f$ ≈ 300 nm, the average $\sigma$ is assessed to be ≈ -3.9 GPa, while $\sigma$ is ≈ -4.2 GPa at $t_f$ ≈ 1615 nm. The $\sigma$ vs. $t_f$ in $Al_{0.67}Ti_{0.33}N$ coating with $V_s$ = -80 V is a bit different from that of other coatings. It appears that when $t_f$ in the range of 300 to 600 nm, the $\sigma$ is increased from -6.4 GPa to -6.9 GPa (see Figure 10 (e)). Beyond $t_f$ ≈ 600 nm, $\sigma$ consistently decreases with the increase in $t_f$. The $Al_{0.67}Ti_{0.33}N$ coating deposited with a $V_s$ =-100 V, the average $\sigma$ remains around -6.8 GPa throughout the $t_f$ (i.e., 600 nm < $t_f$ < 1502 nm).

To quantify the evolution of $\sigma$ with $t_f$, experimentally determined data are fitted with Eqn. 3 as Model 1 in Figure 10. The least-squares optimized coefficients and exponents of the power-law are tabulated in Table 2 . The magnitude of $\sigma_0$ increases consistently with the increase in $V_s$ and changes from tensile to compressive as $V_s$ changes from FP to [-20, -100] V (see Table 2). Unlike $Al_xTi_{1-x}N$:x≤0.5, the least-squares optimized values of c obtained for $Al_{0.67}Ti_{0.33}N$ are positive and remain fixed at 3474.1 GPa nm$^\gamma$ with $\gamma$ = 1.29, 1.35, 1.35, and 1.58 for coatings deposited with $V_s$ = FP, -20 V, -40 V, and -60 V, respectively. As mentioned previously, the evolution of $\sigma$ in $Al_{0.67}Ti_{0.33}N$ coating deposited with $V_s$= -80 V is different, and the optimized $\sigma_0$, c, and $\gamma$ for the coating are -5.8 GPa, -6.1 GPa nm$^\gamma$, and 0.25 respectively (see Figure 10 (e) and Table 2).

It is to note that the estimated $D_{111}^{IP}$ of $Al_{0.67}Ti_{0.33}N$ coatings are almost constant over $t_f$ (see Figure 6 (d)). Thus, in fitting Eqn. 4 to model the evolution of $\sigma$ in $Al_{0.67}Ti_{0.33}N$ we used $D_{111}^{\psi \in [45-55]^\circ}$ as $L(t_f)$ (see Figure S-9 (d) of SI and Eqn. 4). The least squared optimized fitting parameters from Eqn. 4 are tabulated in Table 3. Note that we have fixed $\sigma_C$ = -0.02 GPa and best fitted value of $\sigma_{T,0}$ is = 2.5 GPa at $L_0$=100 nm. The optimized value $\beta D_{eff}$ (50.1 nm$^2$/s) for $V_s$ = FP and it increases with increase in $V_s$ (see Table 3). There is a consistent increase in value of $l$ with an increase in $V_s$ and increases from 0.30 nm [$V_s$ = FP] to 0.8 nm [$V_s$ =-60 V]. Coating deposited with $V_s$ = FP, the optimized A is -12.0 GPa, while B is very close to 0. The optimized magnitude of A [and B] consistently increases with increase in |$V_s$|. The best fitted $D_i$ is obtained as 89.9, 74.8, and 87.6 nm$^2$/s for $V_s$ = -20, -40, and -60 V, respectively.

## V. DISCUSSION

Mechanism of formation of texture or preferred orientation of crystal planes in TMNs polycrystalline films are intensely debated in the literature. Broadly four different mechanisms are suggested: (i) minimization of overall energy [strain energy density, stopping energy, and surface free energy], (ii) anisotropic ion channelling along different crystallographic orientations, (iii) ratio of N to transition metals (TMs) at the surface of the substrate, and (iv) mobility of the impinging atoms on the surface of the growing films [44,58,59]. Rafaja *et al.* [60] further suggested that preferred orientations can be influenced by the experimental geometry too. Hultman *et al.* argued that because of the higher mobility of Ti-adatoms on on c-100 surface than c-111 surface [61] at pre-coalescence stage of islands, Ti adatoms have a higher chance of moving off c-001 faceted islands compared to c-111 faceted islands leading to low growth rate of c-001 faceted islands. But once the islands are coalesced, the effect is reversed, i.e., Ti adatoms will have more time to reside on c-111 surfaces than c-001 surfaces, thus the Ti adatoms capturing probability on c-111 surface is higher than on c-001 surface. Thus, as deposition proceeds c-111 oriented grains grow faster than c-001 oriented grains in a competitive growth mode that leads to c-111 preferred orientations of the films. This explains quite adequately the acquired texture from XRD patterns of TiN, $Al_{0.25}Ti_{0.75}N$, and $Al_{0.50}Ti_{0.50}N$ coatings deposited with $V_s$ = FP, where c-111 is the preferred orientation.

Change in the texture orientation upon application of substrate bias voltage has been intensely debated in the literature. Based on first-principles analysis, Gall *et al.* [62] showed that in presence of lower atomic N concentration, the diffusion length of Ti adatoms on c-001 surfaces are comparatively higher than on c-111 surfaces. While at higher atomic N environment, c-200 surfaces will be covered with TiN$_x$ like clusters, which results into a substantial drop of diffusion length of adatoms. In such growth conditions the preferential growth of c-200 grains take place. This partly explains the preferential orientation of our deposited TiN films with substrate bias, |$V_s$| ≥ 20 V, where intensity of diffracted c-200 peak along the GD increases with an increase in $V_s$. While it is quite evident that in TiN, c-111 peak is always predominant along the GD irrespective of $V_s$, a careful visualization of XRD pattern (see Figure 1) reveals that the intensity of c-200 patterns is very strong close to the GD, but tilted ≈ 18°/22° away for coatings deposited with $V_s$ = -20, -40, and -60 V. We suspect such inclined texture of c-200 oriented grains is due to the geometry of the deposition process, where the incidence angle of the arc evaporated ions is at an angle of 35° to the direction substrate normal [63]. Furthermore, the return of c-111 texture from c-200 texture with higher $V_s$ [$V_s$=-80 V and -100 V] could be the consequence of the recrystallization of c-200 orientated grains in to c-111 oriented grains as a consequent of secondary nucleation [64]. Schell *et al.* suggest that different kinds of defects drive the recrystallization process [65], which agrees with the fact that higher substrate bias causes an increase in incident ion energy that can lead to the formation of high densities of surface defects as well as defects in the bulk.

Kinetics of adatoms in Al incorporated TiN is thoroughly investigated by Hultmann et al. [61] using first-principles calculations. They suggested that Al adatoms have higher mobility on TiN(001) than on TiN(111) surfaces. However, because of a lower diffusion activation barrier of Al in TiN(111) than that of Ti limits the kinetic advantage of Al adatoms in development of c-111 texture. The authors argued that incorporation of Al in the TiN(001) surface creates a trap state for Ti near substitutional $Al_{Ti}$ site and Al at $Al_{Ti}$ site of TiN(111). They proposed that in presence of $Al_{Ti}$ trap states the chemical potential advantage for formation of c-111 texture reduces. This agrees well with our observation for $Al_{0.25}Ti_{0.75}N$ and $Al_{0.50}Ti_{0.50}N$ coatings. For a given $V_s$, diffraction intensity from c-200 crystal plane is higher for $Al_{0.50}Ti_{0.50}N$ than $Al_{0.25}Ti_{0.75}N$ coatings. The continuous reduction in the tilt angle of intense diffraction spots of c-200 crystal planes from the GD with increase in $V_s$ for TiN, $Al_{0.25}Ti_{0.75}N$ and $Al_{0.50}Ti_{0.50}N$ suggests that the kinetics of adatoms can overcome the geometry-introduced constraints in the microstructure evolution.

Presence of both cubic and hexagonal phases in $Al_{0.67}Ti_{0.33}N$ deposited with substrate bias $V_s$ = FP, -20, -40, and -60 V bias are in good agreement with the results of theoretical estimations[36,37], where for x ≈ 0.64 -74 both hexagonal and cubic phases can be formed. Daniel et al. [66] demonstrated that beyond $V_s$ of -50 V, hexagonal phases become the minority and eventually vanish at further higher voltages, which agrees with our experimental observations. The preferred orientation of h-10$\bar{1}$0 crystal planes of $Al_{0.67}Ti_{0.33}N$ deposited at $V_s$ = FP can be understood through considering texture formation in PVD coated polycrystalline *h*-AlN [67]. Jin *et al.*[67] suggested that in *h*-AlN there are two types of Al-N bonds available, namely $B_1$ and $B_2$. While bond $B_1$ is more covalent, $B_2$ is more ionic [67] and the bond energy of type-$B_2$ is smaller than that of $B_1$, thus bond type $B_2$ is relatively easy to break. On the *h*-10$\bar{1}$0 surfaces of *h*-AlN, only $B_1$ type bonds, while on *h*-0001 surfaces both $B_1$ and $B_2$ types are populated. Thus, with high energetics ion beams, the probability of survival of bond type $B_1$ in AlN is higher than that of $B_2$ type, indicating h-0001 faceted grains grow faster with high substrate bias. The first principles estimated surface free energy of *h*-AlN [68] suggests that the *h*-10$\bar{1}$0 surface has lower surface free energy than that of *h*-0001 surfaces, suggesting with non-energetics ion beams the *h*-10$\bar{1}$0 surface will grow along the GD. Cheng *et al.* [69] also suggested that for the development of closed packed surface planes (i.e. h-0001 surfaces) adatoms need to stay a longer time at the surface than that of loosely packed surface (i.e. *h*-10$\bar{1}$0). The preferred orientation of *h*-10$\bar{1}$0 along the GD in $Al_{0.67}Ti_{0.33}N$ coating deposited at $V_s$ = FP agrees with predictions of three different mechanisms as discussed above. With increase in $|V_s|$, together with breaking of $B_2$ type bonds along with the higher diffusivity of adatoms causes preferred orientation of *h*-0002 along the GD. The orientation of c-111 and c-200 crystal planes of c-$Al_{0.67}Ti_{0.33}N$ is a bit different than that of $Al_xTi_{1-x}N$; x ≤ 0.5. The intense spot of c-111 always remains fixed at an angle of 42° to the GD, while for $|V_s| \geq 60$ V, the intense spot in c-200 Debye-Scherrer rings tilted away from 16° to 24° from the GD.

The correlation between ***k*** and ***n***, obtained by fitting the evolution of crystallite size, is estimated based on the Kendall rank correlation test [70]. The estimated *p*-values of ***k*** and ***n*** for individual compositions are .021, .083, .017, and .055 (.007) for TiN, $Al_{0.25}Ti_{0.75}N$, $Al_{0.50}Ti_{0.50}N$, $Al_{0.67}Ti_{0.33}N$ ($Al_{0.67}Ti_{0.33}N$: $\psi \in$ [45°-55°]), respectively. This suggests that a significant correlation between estimated ***k*** and ***n*** exists. We also checked Pearson's correlation between ***k*** and ***n***, where the magnitude of correlation coefficient is higher than 0.88, hinting to a strong linear correlation, indeed we find that there is a strong linear correlation between log(***k***) and ***n*** (see Figure S-9 of SI) [71]. In such case, Depla *et al.*[71] proposed that the power-law [Eqn. 2] need modifications, and the modified version of the power-law is given by;

$$\frac{D_{hkl}}{D_0} = \left(\frac{t_f}{t_0}\right)^{\mathbf{n}} \qquad (5)$$

where $D_0$ and $t_0$ are two constants. Fitting of $D_{111}^{IP}$ vs. $t_f$ with Eqn. 5 revealed that the $D_0$ and $t_0$ values for all $Al_xTi_{1-x}N$ are very close to each other, and the mean values are 13.43 ± 3.60 nm, and 12.41 ± 2.40 nm. Fitted $D_0$ and $t_0$ for all $Al_xTi_{1-x}N$ coatings at various $V_s$ are further tabulated in section SI. IV of SI. The optimized ***n*** values are tabulated in Table 4. The value $t_0$ approximates very close to the mean film thickness for continuous film formation [71–73]. Thus, here we argue that the critical thickness for which the arc evaporated $Al_xTi_{1-x}N$ transitions from discrete islands to continuous films happens to be close 12.41 ± 2.40 nm.

The homologous temperature, defined as the ratio of substrate temperature (~ 450 °C [~723 K]) and melting temperature of materials (~2949 °C [~ 3222 K] [74], of our deposited films are close to 0.22. At such low homologous temperature a substantially lower value of mean growth exponent, ***n*** < 0.35, is often observed[71] in agreement with our estimation. The ***n*** [from Eqn. 2 and Eqn. 5] is higher with films deposited with $V_s$ = FP and lays between ≈ 0.25 to 0.4. With $V_s$, ***n*** reduces to ≈ 0.15 to ≈ 0.18 at -20 V and further increases to ≈ 0.23 at $V_s$ =-100 V (see Table 4). This "U" shape of ***n*** vs $V_s$ suggests a competitive growth mechanism, possibly a competition between nucleation rates caused by energetics ions, creation of defects, and diffusion of energetics adatoms [75]. A consistent increase in average grain size, $D_{111}^{IP}$, with an increase in film thickness is a consequence of grain coarsening, coalescence of smaller grains to larger grains, and grain growth below the surface of growing films, as the film thickness/growth duration increases [76]. The *nanocrystalline*-like surface morphologies (see Figure 3(p-u) and Figure 5(a-1, b-1, and c-1)) and evolution of grain size (see Figure 6 (d) and Figure S-9 (d) of SI) of $Al_{0.67}Ti_{0.33}N$ coatings suggests that a frequent re-nucleation of grains occurs during the entire growth durations. Our TEM analysis indicates that formation of hexagonal phases around the grain boundaries is the reason behind the stagnation and, hence, the frequent re-nucleation of grains.

Table 4: Optimized exponent **n** as obtained by fitting Eqn. 5 with the $D^{IP}_{111}$ vs. $t_f$ of $Al_xTi_{1-x}N$ coatings on Si (100) substrate.

| $V_s$ (V) | $n$ (TiN) | $n$ ($Al_{0.25}Ti_{0.75}N$) | $n$ ($Al_{0.50}Ti_{0.50}N$) | $n$ ($Al_{0.67}Ti_{0.33}N$) |
|---|---|---|---|---|
| FP | 0.25 | 0.33 | 0.40 | -0.09 |
| -20 | 0.15 | 0.12 | 0.09 | 0.01 |
| -40 | 0.18 | 0.14 | 0.18 | -0.01 |
| -60 | 0.19 | 0.14 | 0.15 | 0.04 |
| -80 | 0.23 | - | 0.15 | 0.09 |
| -100 | 0.23 | - | - | -0.10 |

The generation of compressive stress in PVD coated thin films is often observed. Among the many mechanisms, the insertion of atoms into the grain boundary has been gaining more credibility [56,77]. However, in the deposition of coatings with high energetics vapour flux, stress can be caused by additional factors, too. A comprehensive model proposed by Chason *et al.*[56] includes such factors [ e.g., boundary coalescence, formation of the new grain boundaries, and their densification and creation of defects in bulk of films].

In general, it is seen here that the $\beta D_{eff}$, which is a measure of effective diffusivity, increases with increase in $V_s$. However, in our TiN coatings $\beta D_{eff}$ shows a "U" type behaviour with increase in $V_s$. This behaviour coincides with the "U" type character of the evolution of the texture intensity ratio $I_{c-200}/I_{c-111}$ (see Figure S-5 of SI). The $\beta D_{eff}$ in $Al_{0.25}Ti_{0.75}N$, and $Al_{0.50}Ti_{0.50}N$ coatings deposited with $V_s$ = FP is substantially lower than for coatings deposited with applied negative $V_s$ [except for coatings with cohesive failure is seen]. This applied $V_s$ influences diffusivity of adatoms, and it increases with an increase in ion energies. This agrees with the fact that with increase in $|V_s|$ wider nanocolumnar dense coatings are seen. The $\beta D_{eff}$ of TiN, $Al_{0.25}Ti_{0.75}N$, and $Al_{0.50}Ti_{0.50}N$ is within the range of 47-210 nm$^2$/s, much higher than that reported for different metals and TiN thin films deposited by magnetron sputtering ($\beta D_{eff}$ = 0.12 nm$^2$/s for TiN) [30,78]. For "non-energetics" electrodeposited Ni and Cu samples the reported absolute values of $\beta D_{eff}$ varied quite differently from that mentioned earlier. Chason *et al.* [79–81] reported $\beta D_{eff}$ to be around 174 nm$^2$/s to 1761 nm$^2$/s in the same order as our predictions. The fitted parameters A and B, for arc evaporated TiN, that we obtained from least-squares fitting, are in agreement with different materials that are discussed in the literature [30,56,78]. The value of $l$ increases with an increase in $V_s$ further agreeing that with an increase in kinetic energy, ions can penetrate more into the bulk of thin films from the surfaces. The obtained order of $l$ agrees with the ranges of the value predicted [30]. We find that the evolution of A and B with $V_s$ is not necessarily linear. In TiN coatings, the magnitude of final stress level at the end of deposition is of the "U" type with the $V_s$. This suggests that the A and B in the kinetic model may have a dependence on the texture/microstructure of the polycrystalline thin films and do not necessarily increase with an increase in the ion energies alone. Interestingly, the evolution of A and B obtained from the least square fitting of stress evolution of $Al_{0.25}Ti_{0.75}N$ and $Al_{0.50}Ti_{0.50}N$ with different $V_s$ are a bit different from TiN. In the case of $Al_{0.25}Ti_{0.75}N$, values of both A and B increases with an increase in $V_s$, except for $V_s$=-60 V where a small decrease in A is noticed.

No strong power-law type evolution of σ is observed in $Al_{0.50}Ti_{0.50}N$ deposited with $V_s$ = -80 V, suggesting high ion energies can create a large density of defects, including dislocations and twin boundaries [82,83], which can give rise to high stress state in the coating. At the same time high densities of surface defects can hinder the mobility of adatoms, which adds to the already high σ to the films. This hypothesis is further tested for $Al_{0.25}Ti_{0.75}N$ and $Al_{0.50}Ti_{0.50}N$ coatings deposited at $V_s$ =-60 V, where the cohesive failure of films is seen. The $\beta D_{eff}$ values for both these coatings are substantially lower than for the coatings deposited at $V_s$ = -40 V (see Table 3) suggesting low diffusivity of adatoms on surface may be one of the reasons for higher internal stress in the coating.

The estimated values of diffusivity of defects, $D_i$, generated in the bulk part of the materials are diverse in the literature. For sputter-deposited TiN, Mo, and Cu, the $D_i$ is estimated, using the kinetic model as discussed in section III, around 9.87 nm$^2$/s, 0.28 nm$^2$/s and 5503.14 nm$^2$/s [30,56,78]. Through numerical simulations, Pasianot *et al.* [84] and Starikov *et al.* [85] predict the value of $D_i$, ~ 1260 nm$^2$/s and 10$^{10}$ nm$^2$/s respectively. Our predicted diffusivity $D_i$, with TiN, $Al_{0.25}Ti_{0.75}N$, and $Al_{0.50}Ti_{0.50}N$ are in the order of $10^1$-$10^2$ nm$^2$/s. At this stage, we do not find any correlation between the $V_s$ and $D_i$. Since no reports are available regarding the estimation of $D_i$ for arc evaporated coatings in the literature, we could not compare our estimates either. The cohesive failures in PVD coated films are well documented and are often attributed to high stress level, i.e., when the stress level is higher than the fracture toughness, the coating releases the excess strain energy through the formation of cracks in the coating [47].

The evolution of σ in Al$_{0.67}$Ti$_{0.33}$N coatings is more scattered in comparison to Al$_x$Ti$_{1-x}$N; x ≤ 0.5. This is a consequence of a larger stress distribution caused by finer grains, compositional fluctuations, and formation of multi-phase crystal structure etc. High tensile stress in Al$_{0.67}$Ti$_{0.33}$N deposited at $V_s$ = FP is possibly the consequence of very small grain size (L) as tensile stress, $\sigma_T$, scales as $L^{-0.5}$. Through a detailed molecular dynamics simulation of bcc metals, Thompson *et al.* [86] proposed that stress evolution is also dependent on surface morphology. They found that when grains are well separated, and impinging energy of atoms is lower (~ 5 eV) the stress is tensile, and it scales with reduced contact area between the islands. Cross-sectional SEM and TEM suggest a "*perpendicular edge*" like structure, indicating that the surface morphology is the key behind this tensile stress. A flatter surface morphology in Al$_{0.67}$Ti$_{0.33}$N coatings deposited with different $V_s$ along with the mechanisms of Eqn. 4 is responsible for compressive stress. Unlike Al$_x$Ti$_{1-x}$N; x≤0.5, in Al$_{0.67}$Ti$_{0.33}$N stress increases with increase in $t_f$ ($V_s$ = -20, -40, and -60 V). The least-squared fitted c for Model 1 is the same for Al$_{0.67}$Ti$_{0.33}$N coatings for $V_s$ = FP, -20, -40, and -60 V, which is probably the consequence of similar crystallite sizes ($D_{111}^{IP}$ ≈ 8-12 nm). The origin of the unusually small, but finite decrease in σ with the increase in $t_f$ in Al$_{0.67}$Ti$_{0.33}$N coatings deposited with $V_s$ = -80 V is unknown to us. We suggest such behaviour might have a pure microstructure origin or is the consequence of the formation of cubic phase crystal only. With $V_s$ = -100, the estimated σ is almost constant, suggesting that formation of large densities of defects due to energetic ions are dictating the stress levels in the coatings. The fitted value of $\beta D_{eff}$ increases with an increase in $V_s$ from FP to -60 V suggesting the net diffusivity of adatoms increases despite frequent re-nucleation. The value of A, B, and $l$ increases with an increase in $V_s$, which agrees with the suggestions made by Chason *et al.* [56]. The best-fitted value of B from the Al$_{0.67}$Ti$_{0.33}$N coating deposited with $V_s$ = FP is obtained as 0, as at no $t_f$, σ is compressive. A value of $D_i$ ≈ 75-90 nm$^2$/s is noted. As already discussed earlier, there are no reports in the literature for which $D_i$ values can be compared for arc evaporated films.

In short, an effective compressive stress relaxation in Al$_x$Ti$_{1-x}$N; x≤0.5 coatings with the increase in $t_f$ is seen, and it is a consequence of the average increase in crystallite size (also grain size) at higher $t_f$. An accumulation of compressive stress is observed in Al$_{0.67}$Ti$_{0.33}$N, where a substantial hexagonal phase is recorded. In absence of hexagonal phases, a small but noticeable compressive stress relaxation is seen. Thus, here we argue that stress relaxation in Al$_x$Ti$_{1-x}$N coatings with increase in film thickness is a combined effect of grain growth as well as the fraction of hexagonal phases in them.

*Table 5: Critical stress ($\sigma_{cr.}$), critical film thickness ($h_{cr.}$), plain strain fracture toughness ($K_{IC}$), and elastically stored energy ($G_s$) obtained for Al$_{0.25}$Ti$_{0.75}$N deposited with substrate bias ($V_s$) of -60 V, and Al$_{0.50}$Ti$_{0.50}$N deposited with substrate bias ($V_s$) of -60 V, and -80 V.*

| Composition | $V_s$ (V) | $\sigma_{cr.}$ (GPa) | $h_{cr.}$ (nm) | $K_{IC}$ (MPa.m$^{1/2}$) | $G_s$ (J/m$^2$) |
|---|---|---|---|---|---|
| **Al$_{0.25}$Ti$_{0.75}$N** | -60 | -4.9 ± 0.10 | 770 ± 20 | 3.09 ± 0.05 | 23.5 |
| **Al$_{0.50}$Ti$_{0.50}$N** | -60 | -4.4 ± 0.15 | 950 ± 50 | 3.04 ± 0.12 | 23.8 |
| **Al$_{0.50}$Ti$_{0.50}$N** | -80 | -7.9 ± 0.18 | 745 ± 25 | 4.84 ± 0.12 | 55.9 |

As mentioned earlier, the cohesive failure of the coatings is attributed to a high-stress field in the coatings, and the critical stress at which it fails is characteristic of fracture toughness of the materials. In the thin film forms, it is difficult to estimate the fracture toughness due to their small size and unreliable methods. Huang *et al.* [47] developed a method by using stored internal energy. In that proposed method, many polycrystalline thin film coatings were deposited with varying thickness and residual stress (σ). The shortcoming of their work is the exact determination of the film thickness, at which the cracks form and propagate. This shortcoming can be overcome by *in-situ* measurements, such as utilized in the current work. Since from *in-situ* XRD data analysis, we can determine critical residual stress ($\sigma_{cr.}$) and critical film thickness ($h_{cr.}$) at which initiations and propagations of cracks occur [i.e., location of discontinuity in XRD peak intensity, crystallite size and stress evolution]. For thin films, Huang *et al.* [47] approximated the plain strain fracture toughness ($K_{IC}$) to elastically stored energy ($G_s$) and critical stress and $h_{cr.}$ by the relation given below.

$$K_{IC} = \frac{1}{\sqrt{2}} |\sigma_{cr.}| \sqrt{h_{cr.}} = \sqrt{\frac{E_{hkl} G_s}{(1 - \nu_{hkl}^2)}} \qquad (6)$$

Utilizing the values of σ and $t_f$, at which discontinuity in stress and diffraction intensity evolution happens and using them as $σ_{cr.}$ and $h_{cr}$ respectively in Eqn. (6), we estimate the $K_{IC}$ and $G_s$ for samples, where cohesive failure is seen. The values are tabulated in Table 5.

The estimated $K_{IC}$ values closely agree with the estimate of Mayrhofer *et al.* ($K_{IC}$ = 2.7 ± 0.3 MPa.m$^{1/2}$ for $Al_{0.60}Ti_{0.40}N$ [87] and 3.5 ± 0.3 MPa.m$^{1/2}$ for $Al_{0.46}Ti_{0.54}N$ [88]). The $K_{IC}$ estimated for $Al_{0.50}Ti_{0.50}N$ with $V_s$ = -80 V is a bit higher. A relatively higher $K_{IC}$ of $Al_{0.25}Ti_{0.75}N$ and $Al_{0.50}Ti_{0.50}N$ than that of TiN [47,89] agrees with the theoretical prediction of Abrikosov *et al.*[90].

## VI. SUMMARY

In summary, we investigate the real-time evolution of stress and microstructure of aluminium titanium nitride ($Al_xTi_{1-x}N$, 0.0 ≤ x ≤ 0.67) thin films on Si (100) substrate during their deposition with a custom-designed cathodic arc deposition technique. The synchrotron-based 2-dimensional in-situ x-ray diffractograms of coatings are recorded during the depositions. The strain and biaxial stress are evaluated by utilizing $sin^2ψ$ method, and their evolution with film thicknesses is analysed within a kinetic model. Line profile analysis is carried out to quantify the evolution of crystallite sizes and texture formation. Thorough ex-situ characterizations of the coatings were carried out using electron microscopies tools. The main conclusions derived from this work are as follows.

- While in coatings of $Al_xTi_{1-x}N$: x ≤ 0.5 long columnar morphologies are formed, where the domain size of nanocolumns and the density of coatings increase with an increase in substrate bias voltage. Owing to higher adatoms' effective diffusivity, $Al_{0.67}Ti_{0.33}N$ coatings have nanocrystalline-like surface morphology. The nanocrystalline-like surface morphology is driven by the re-nucleation of grains and is a consequence of the formation of mixed hexagonal and cubic phases.
- At lower energy of incomings ions (i.e., substrate at floating potential), 111 crystal planes of cubic phases of $Al_xTi_{1-x}N$: x ≤0.5 are preferentially oriented along the growth direction while in $Al_{0.67}Ti_{0.33}N$ coating 220 crystal planes of cubic phases and $10\bar{1}0$ and $11\bar{2}0$ crystal planes of hexagonal phases are oriented parallel to the interface of coatings and substrates. Irrespective of Al content in $Al_xTi_{1-x}N$ coatings, an increase in ion energies drives 200 crystal planes of cubic phases to preferentially orient along/close the GD.
- A power law-like evolution of crystallites' dimension along the in-plane direction with film thickness is observed. While for $Al_xTi_{1-x}N$:x≤0.5 coatings the exponent of the power is positive and lies between 0.15 and 0.4, the exponent for $Al_{0.67}Ti_{0.33}N$ is within the range of ± 0.1. Analysis of the evolution of crystallite size vs. coating thickness revealed that in arc evaporated $Al_xTi_{1-x}N$ coatings the critical thickness of transformation from discrete islands to continuous films is 12.41 ± 2.40 nm.
- The biaxial stress in $Al_xTi_{1-x}N$ coatings follows a power law-like behaviour with thickness of thin film. The converged value of stress in $Al_xTi_{1-x}N$ coatings deposited at floating potential substrate bias is always tensile, while they are compressive with the negative substrate bias voltages. In all $Al_xTi_{1-x}N$: x ≤0.5 coatings a biaxial stress relaxation with increase in film thickness is observed, while in $Al_{0.67}Ti_{0.33}N$ coatings an increase in biaxial compressive stress with film thickness is observed. Kinetic modelling analysis suggests that upon a negative substrate bias a two-to-three-fold increase in effective diffusivity of adatoms occurs. The length ion induced densification of grain boundaries and to the depth at which defects are being created increases from 0.2-0.3 nm to 0.7-0.8 nm, as the substrate bias changes from the floating potential to –100 V.
- Based on their critical stress level at the critical film thickness before fracture, the estimated fracture toughness of $Al_{0.25}Ti_{0.75}N$ and $Al_{0.50}Ti_{0.50}N$ coatings are in the 3.09 ± 0.1 MPa.m$^{1/2}$ and [3.04 ± 0.1, 4.84 ± 0.13] MPa.m$^{1/2}$ respectively agreed well with estimation from other techniques reported in the literature.

**Acknowledgements:** All authors acknowledge DESY for beamtime under proposal I-20210060EC. We acknowledge financial support from the Swedish Research Council via the Röntgen Ångström Cluster (RÅC) Frame Program.